\documentclass[twocolumn]{aastex631}

\newcommand{\be}{\begin{equation}}
\newcommand{\ee}{\end{equation}}

\def\nova{V1674 Her}

\begin{document}

\title{The Rise of Nova V1674 Herculis}

\author[0000-0001-9171-5236]{Robert M. Quimby}
\affiliation{Department of Astronomy and Mount Laguna Observatory\\
San Diego State University, San Diego, CA 92182, USA}
\affiliation{Kavli Institute for the Physics and Mathematics of the Universe (WPI)\\ The University of Tokyo Institutes for Advanced Study\\
The University of Tokyo, Kashiwa, Chiba 277-8583, Japan
}

\author[0000-0002-4670-7509]{Brian D. Metzger}
\affiliation{Department of Physics and Columbia Astrophysics Laboratory, Columbia University, New York, NY 10027, USA\\}
\affiliation{Center for Computational Astrophysics, Flatiron Institute, 162 5th Ave, New York, NY 10010, USA\\}

\author[0000-0002-9632-6106]{Ken J. Shen}
\affiliation{Department of Astronomy and Theoretical Astrophysics Center, University of California, Berkeley, CA 94720, USA\\}

\author[0000-0002-1276-1486]{Allen W. Shafter}
\affiliation{Department of Astronomy and Mount Laguna Observatory\\
San Diego State University, San Diego, CA 92182, USA}

\author[0000-0002-6339-6706]{Hank Corbett}
\affiliation{Department of Physics \& Astronomy\\
University of North Carolina at Chapel Hill, Chapel Hill, NC 27514, USA}

\author{Madeline Overton}
\affiliation{Department of Physics and Astronomy and Nevada Center for Astrophysics, University of Nevada, Las Vegas, 4505 South Maryland Parkway, Las Vegas, NV 89154, USA}

\begin{abstract}

Observational constraints on classical novae are heavily biased to phases near optical peak and later because of the simple fact that novae are not typically discovered until they become bright. The earliest phases of brightening, coming before discovery, are typically missed, but this is changing with the proliferation of wide-field optical monitoring systems including ZTF, ASAS-SN, and Evryscope. Here, we report on unprecedented observations of the fast nova \nova\ beginning $>$10 mag below its optical peak and including high-cadence (2\,min.) observations that chart a rise of $\sim$8\,mag in just 5\,hours. Two clear breaks are identified as the light curve transitions first from rising slowly to rising rapidly, followed by a transition to an even faster, nearly linear rate of increasing flux with time. The depths of the observations allow us to place tight constraints on the size of the photosphere under the assumption of blackbody emission from a white dwarf emitting at its Eddington luminosity. We find that the white dwarf was unlikely to have overflowed its Roche lobe prior to the launch of a fast wind, which poses a challenge for explaining the Fermi $\gamma$-ray detections as the interaction of a fast wind with a slow-torus of gas stripped from the inflated white dwarf envelope by the companion. High-cadence observations of novae from Evryscope and the planned Argus Array can record the diversity of rising nova light curves and help resolve how the interplay between thermonuclear fusion, binary interaction, and shocks power their earliest light.

\end{abstract}


\section{Introduction} \label{sec:intro}

Classical novae, non-terminal outbursts that occur when the envelope accreted onto a white dwarf (WD) star ignites (see \citealt{cms2021} for a review), are typically discovered near peak optical brightness\footnote{For a listing of nova discovery and peak magnitudes, see \url{https://github.com/Bill-Gray/galnovae/blob/master/galnovae.txt}}. Follow-up campaigns can be initiated soon thereafter to record the evolution of the nova in detail as it fades, but the rising portion of the light curve is typically poorly constrained, especially for the most rapidly evolving novae (e.g., see examples in \citealt{2010AJ....140...34S}).

In most cases the nova is first detected when it is within 2 mag of its optical peak, which, by definition of a nova, means that at least the first 6 mag of brightening are missed. Some novae have historical imaging that provides detections of the progenitor systems in quiescence prior to outburst \citep{schaefer_AAS2021}, but almost all lack a detailed record of how the nova transitioned from its quiescent state to outburst. Notable exceptions include observations of RS Oph, V598 Pup, and KT Eri starting $\sim$4\,mag below peak with the Solar Mass Ejection Imager \citep{hou10}, observations of the very slow Nova Velorum 2022 (Gaia22alz) starting $\sim$7\,mag below peak \citep{aydi2023}, observations of ASASSN-16ma starting $\sim$8\,mag below peak \citep{Li2017}, and observations of T Pyxidis, whose 2011 outburst was discovered and then intensively followed up starting $\sim$8\,mag below peak \citep{schaefer2013}. In this study we discuss the unprecedented $>$10\,mag rise of Nova Herculis 2021 (\nova) and how its early light curve can be used to constrain the initial expansion phases of the outburst.

Novae were once thought to be powered exclusively by the radiation emitted by thermonuclear fusion at the base of the accreted envelope \citep{Gallagher&Starrfield78}, but the shocking discovery of $\gamma$-ray novae has forced a revision of the theory \citep{ackermann2014}. Observational evidence still shows that nova systems contain a white dwarf interacting with its companion \citep{Patterson84}, and thus it is believed that stable accretion onto the WD builds an envelope of fuel until the rate of energy generated by thermonuclear fusion exceeds the rate this energy can diffuse away (e.g., \citealt{Prialnik+79}). This is often described as ``thermonuclear runaway'' in the literature, but we emphasize that there can be a long simmering phase prior to the convective zone reaching the photosphere. It is at this point the luminosity of the WD photosphere begins to increase, but initially the radius of the WD remains essentially constant. As the luminosity approaches the Eddington limit, the WD envelope begins to expand and the temperature drops accordingly to maintain the Eddington luminosity. The increasing radius and lowering temperature conspire to produce a rise in optical flux. When the photospheric temperature drops to roughly $1.5 \times 10^{5}$\,K, the opacity increases due to Fe-peak elements, and a fast, optically thick wind may be launched \citep{kato_hachisu1994}. However, this temperature transition may occur at around the time the WD fills its own Roche lobe, which means the companion may be important to what follows \citep{shen_quataert2022}.

One possibility is that once the WD fills its Roche lobe material begins flowing to the companion, which, having already filled its own Roche lobe, cannot accrete further material. Thus the extra material ends up ejected from the L2 nozzle opposite the companion from the WD and a torus of gas forms in the equatorial plane of the system (e.g., \citealt{pejcha2016,lu2023}). This scenario is supported by high-resolution radio imaging (e.g., \citealt{chomiuk2014}) and spectroscopic observations of nova outbursts that show P-Cygni profiles prior to maximum light with absorption troughs blue shifted by velocities consistent with the orbital velocities of the binary \citep{aydi2020b}. When the fast wind is launched, it may interact with this relatively slow-moving torus to produce $\gamma$-ray emission as well as optical light \citep{Metzger+14,Li2017, munrai2017, aydi2020}. Nova spectra taken after maximum light have shown multiple velocity components that can be identified as the slow torus, the fast wind, and intermediate-width features that can be identified as the interaction region of these two.

A white dwarf filling its Roche lobe and emitting at its Eddington luminosity will have an absolute $g$-band magnitude in the optical of $M_g \sim -1$, which is $\sim$6\,mag below the typical peak of novae. As mentioned, novae are not typically discovered until they are much brighter in the optical, so there is a lack of direct observational data documenting potential interactions between the WD and its companion at this phase. The rise of wide field optical transient surveys is, however, changing this picture. The Zwicky Transient Facility (ZTF; \citealt{bellm2019, graham2019}), the All-Sky Automated Survey for Supernovae (ASAS-SN; \citealt{Shappee+14, Kochanek+17},
and others now routinely capture snapshots of large swaths of sky each night. These surveys do not always first identify novae at young phases because of the candidate vetting process, but the images are archived and can be mined after discovery. Evryscope takes wide field monitoring a step further by obtaining not just snap shots but nightly movies built from hundreds of exposures covering thousands of square degrees simultaneously \citep{law14}. These surveys operate at depths sufficient to detect the earliest evolutionary phases of galactic novae.

\nova\ has been the subject of multiple studies (e.g., \citealt{woodward21, Drake+21, Patterson+22, sokolovsky2023, Bhargava+24, habtie2024}) due to several unique factors: 
1) with a peak apparent magnitude of $V\sim 6.2$ (see Appendix~\ref{sec:t2}), it is among the top 5\% brightest novae found this century;
2) the visual decline of 2 magnitudes from peak in just 1 day makes it perhaps the fastest novae on record;
3) extensive pre-outburst imaging of the progenitor system by ZTF place tight constraints on the white dwarf spin period prior to outburst \citep{mro21};
4) Fermi detected a coincident $\gamma$-ray source that turned on during the optical peak \citep{sokolovsky2023}; and
5) periodicity was detected in the X-Ray and optical light curves just two weeks after the optical peak that reveal the orbital period \citep{Patterson+22}.
In this paper we focus on another remarkable aspect of \nova: pre-discovery images from ASAS-SN and Evryscope that begin with the nova 10 magnitudes below its optical peak and capture a gradual brightening phase, a quick transition to a fast rise, and a second transition to an even faster rise.

We discuss some of the key observational constraints of the \nova\ system in \S\ref{sec:nova} and provide estimates for the scale of the system in \S\ref{sec:physical}. In \S\ref{sec:phot} we present new analysis of pre-discovery optical photometry from ASAS-SN and Evryscope including discussion of a process we dub ``image model subtraction'' that we use to extract photometry from the crowded Evryscope images (\S\ref{sec:imsub}). We analyze the rising light curve in \S\ref{sec:lc}, and construct simple models in \S\ref{sec:models} to help interpret these data. Discussion and conclusions are offered in \S\ref{sec:conclusions}. For convenience, we adopt an agnostic reference epoch, $t_{\rm ref}$, of HJD 2459377.5 (2021 July 12).

\section{V1674 Her} \label{sec:nova}

Seiji Ueda of Japan discovered \nova\ at $t_{\rm ref} + 0.5411$\,d and promptly posted the discovery to the Central Bureau for Astronomical Telegrams\footnote{\url{http://www.cbat.eps.harvard.edu/unconf/followups/J18573095+1653396.html}} (CBET 4976).
The transient coincides with Gaia DR3 4514092717838547584 at $\alpha$ = 18:57:30.98324, $\delta$ = +16:53:39.5895, which has a pre-outburst Gaia magintude of $G = 19.950 \pm 0.020$. The coordinates also match a PanSTARRS source with $g_{\rm ps} = 20.476 \pm 0.071$, $i_{\rm ps} = 19.483 \pm 0.024$, and $z_{\rm ps} = 19.287 \pm 0.004$. Photometry in the PanSTARRS $r_{\rm ps}$ and $y_{\rm ps}$ bands is not given in the DR2 release; however, the source is clearly detected in stacked images in these bands as well. 

\begin{figure*}
    \centering
    \includegraphics[width=2\columnwidth]{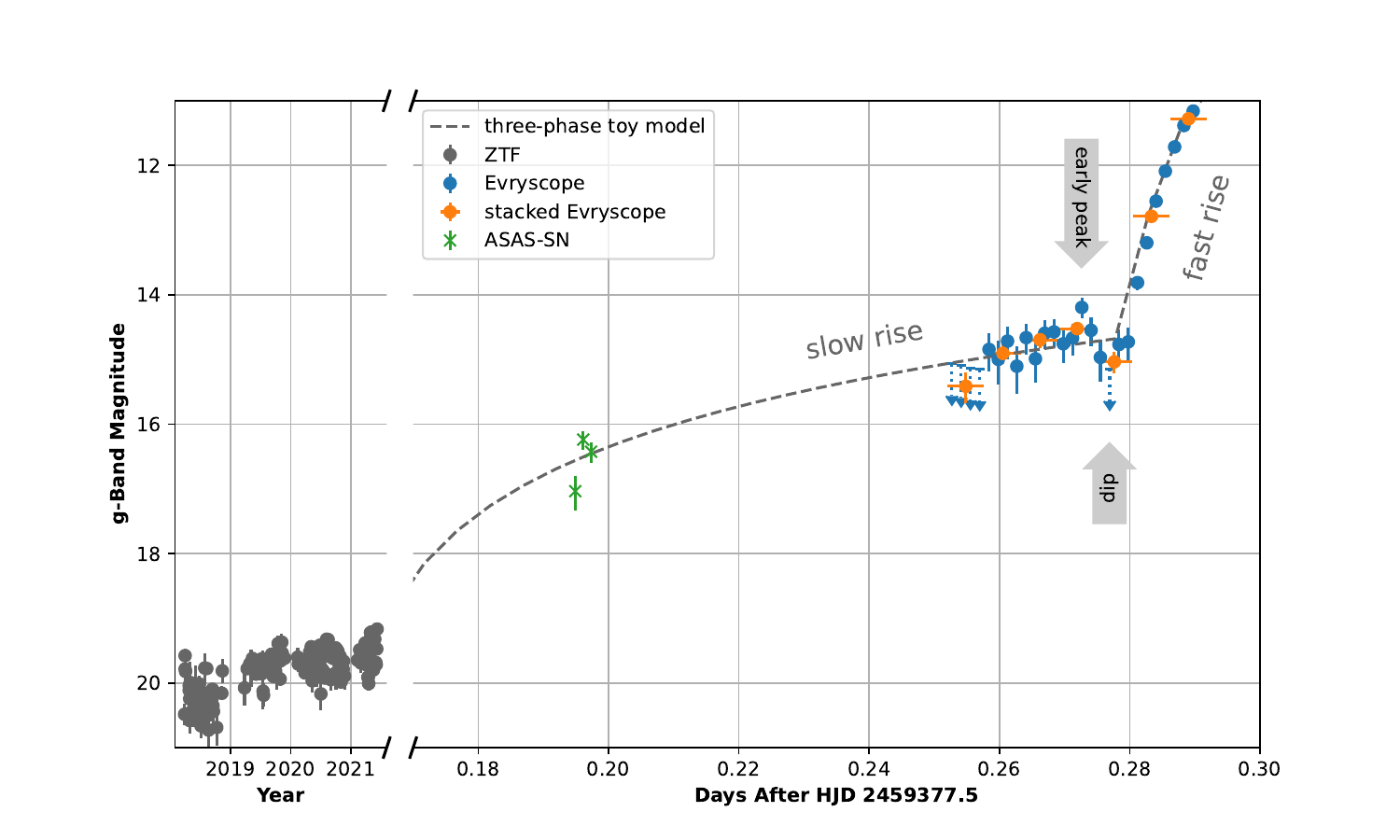}
    \caption{Historical ZTF $g$-band light curve (grey dots) compared to the earliest ASAS-SN (green Xs) and Evryscope (blue dots) detections. The orange dots are the magnitudes after stacking 4 consecutive Evryscope observations. The dashed line shows the best-fit three-phase toy model (see \S\ref{sec:toy_model3}). The plot is truncated well below \nova's visual peak of 6.2\,mag at about HJD 2459377.5 $+0.84$\,days to emphasize the early evolution.}
    \label{fig:early_lc}
\end{figure*}

\begin{figure}
    \centering
    \includegraphics[width=\columnwidth]{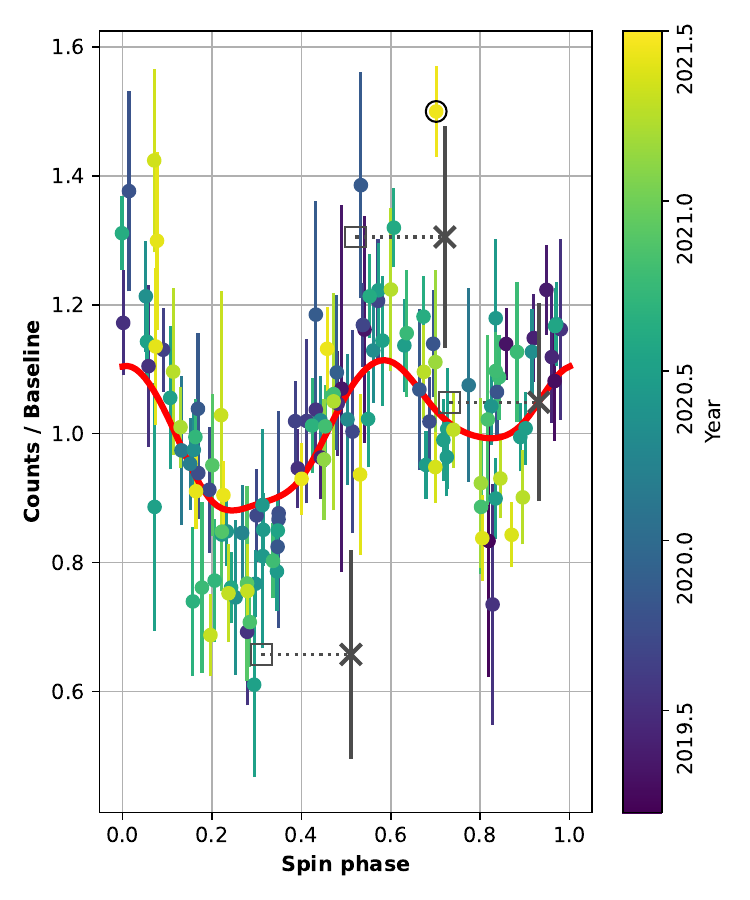}
    \caption{Ratio of the ZTF $g$-band flux to a best-fit 4th order polynomial fit from 2019 through $t_{\rm ref}$ folded by the 501\,s spin period (colored dots). The final, anomalously bright measurement prior to $t_{\rm ref}$ is circled. The red curve shows a spline fit to the better-sampled but lower-amplitude ZTF $r$-band light curve for comparison. The grey crosses show the first three ASAS-SN detections divided by the best-fit three-phase toy model from \S\ref{sec:toy_model3}, and the grey squares mark these same values shifted by $0.2 P_{\rm wd}$. }
    \label{fig:spin_period}
\end{figure}

\nova\ was also detected by the ZTF survey prior to outburst. We use the ZTF forced-photometry service \cite{masci2023} to extract the historical $g_{\rm ztf}$ (Fig.~\ref{fig:early_lc}) and $r_{\rm ztf}$ band photometry. 
In 2019, the average brightness was $g_{\rm ztf} = 20.27$,  $r_{\rm ztf} = 18.81$, but the progenitor was brighter and bluer the following year at $g_{\rm ztf} = 19.74$, $r_{\rm ztf} = 18.59$, and it remained at this higher level until the 2021 nova eruption. The ZTF observations also reveal significant scatter ($\sigma_g=0.21$\,mag, $\sigma_r=0.10$\,mag), indicating the progenitor is variable on shorter timescales. Periodicity in these variations reveals the spin period of the white dwarf star prior to eruption, $P_{\rm wd} = 501.4277 \pm 0.0004$\,s \citep{mro21, Drake+21}. At $g_{\rm ztf} = 19.17$, the ZTF observation on $t_{\rm ref} - 1.5746$\,d is significantly brighter than expected from this trend, and may suggest either flickering or a more steady increase in brightness leading up to the nova outburst (see Fig.~\ref{fig:spin_period}). 

The progenitor system is classified as an intermediate polar in which the white dwarf's strong magnetic field disrupts the inner accretion disk and funnels the accreting matter onto the polar regions \citep{patterson94, Patterson+22}. A hot spot forms on the white dwarf's surface as material accretes, and the intense radiation from this spot modulates as the white dwarf rotates, which results in the observed, short period oscillations in the optical \citep{mro21}. 

Around two weeks after discovery a second periodic signal, $P_{\rm orb}=0.152934 \pm 0.000034$\,d, associated with the orbital period of the binary was also detected in the optical light curve \citep{schmidt2021, shugarov21, Patterson+22}. The orbital period is typical of novae \citep{Schaefer22b}, and the implied binary separation ($\sim$1\,$R_{\odot}$) necessitates a dwarf companion. Using the relation in \citet{Knigge06} the orbital period implies a companion mass of 0.26\,$M_{\odot}$. 

Optical spectra of \nova\ were first obtained at $t_{\rm ref} + 0.84$\,d \citep{mun21}, which is roughly coincident with maximum light in the visual, and $t_{\rm ref} + 0.97$\,d \citep{Aydi+21}. These first spectra were reported to show broad P-Cygni absorption troughs with minima blue shifted by $\sim$3000\,km\,s$^{-1}$ (see also \citealt{habtie2024}). By $t_{\rm ref} + 1.84$\,d the breadth of the lines is reported to have increased to $\sim$5000\,km\,s$^{-1}$. The early spectra are consistent with the \ion{Fe}{2} class but with atypically broad lines \citep{woodward2021ATel}. Later spectra indicate that \nova\ entered the nebular phase by $t_{\rm ref} + 18.4$\,d at which point it began to exhibit strong neon lines \citep{wag21}. This implies a ONe WD, and thus an initial WD mass, $M_{\rm WD} \gtrsim 1.06\,M_{\odot}$ \citep{doherty2015, Drake+21}.

The distance to \nova\ is not well constrained. \citet{schaefer2022} determine a distance range of
2.5\,kpc to 5.4\,kpc (68\% confidence) using a Bayesian calculation that includes a distance likelihood based on the Gaia DR3 parallax (which peaks at negative values for \nova) and priors based on distributions of nova distances determined through other methods. \citet{sokolovsky2023} favor a larger distance of 6.3\,kpc based in part on a lower limit implied by the optical extinction and 3D dust maps. Considering the observed period and absolute Gaia G magnitude ($M_G$) distribution of intermediate polars presented in \citet{mukai2023} (their Fig.~1), we would expect $M_G$ in the 4.2\,mag to 6.2\,mag range. Given the observed Gaia magnitude in quiescence, this corresponds to a distance range of 2.6\,kpc to 6.5\,kpc assuming 1.7\,mag of absorption in the Gaia band.

Fermi LAT detected $0.1-100$\,GeV emission coincident with the location of \nova\ beginning prior to the optical peak \citep{sokolovsky2023}. The $\gamma$-ray source was not present prior to the optical outburst and it faded away in about 18\,h.

Based on \ion{K}{1}\,7699 absorption, \citet{mun21} estimate a reddening of $E(B-V) = 0.55$\,mag. \citet{woodward21} find a consistent reddening value using the diffuse interstellar band at 6613\,\AA. Assuming $R_V = 3.1$, this translates into a $V$-band absorption of $A_V = 1.7$\,mag, and for all $g$-band observations discussed in this paper we assume $A_g = 1.9$\,mag.

\subsection{Estimates of Key Physical Parameters}\label{sec:physical}

Based on the observational constraints (e.g., \citealt{Drake+21,Patterson+22}), we assign a uniform probability for the WD mass in the range $1.0 M_{\odot} \leq M_{\rm WD} \leq 1.35 M_{\odot}$, and we represent the uncertainty in the companion mass with a Gaussian distribution centered at $M_{2} = 0.26 \pm 0.05\,M_{\odot}$. The mass ratio, $q \equiv M_2/M_{\rm WD}$ is then $q=0.22 \pm 0.05$, where the uncertainty covers the 68\% confidence interval. Following \citet{eggleton1983} we estimate a volume equivalent radius for the WD's Roche lobe of $R_{\rm Roche} \sim (4.8 \pm 0.2) \times 10^{10}$\,cm. As the WD expands it may enter a Roche lobe overflow (RLOF) phase where material will flow through the L1 point at $R_{\rm L1}=(3.0 \pm 0.3) \times 10^{10}$\,cm (measured from the center of the WD) and out of the L2 point opposite the companion star at $R_{\rm L2}=(1.90 \pm 0.07) \times 10^{11}$\,cm \citep{lu2023, shen_quataert2022}.

The outer radius of the accretion disk, $R_{\rm out}$, is determined by the circularization radius of matter flowing through the inner Lagrange point (e.g., \citealt{Frank+02}; their Eq. 4.21),
\begin{eqnarray}
R_{\rm out} &\simeq& 4P_{\rm d}^{2/3}(1+q)^{4/3}\left[0.5 - 0.23\,{\rm log}q\right]^{4} R_{\odot} 
\label{eq:Rout}
\end{eqnarray}
where $P_{\rm d} = (P_{\rm orb} / 1\,{\rm day})$. For our adopted WD and companion mass uncertainties this results in $R_{\rm out} = (1.9 \pm 0.1) \times 10^{10}$\,cm.

The inner edge of the disk is determined by the Alfv\'{e}n radius, interior to which the ram pressure of the inflowing gas equals the pressure of the WD magnetic field (e.g., \citealt{Frank+02}; their Eq. 6.18). We have
\be
R_{\rm in} \simeq 1.4\times 10^{9}\,{\rm cm}\,\left(\frac{M_{\rm WD}}{M_{\odot}}\right)^{-\frac{1}{7}}\left(\frac{\dot{M}}{10^{-9}M_{\odot}\,{\rm yr}^{-1}}\right)^{-\frac{2}{7}}\left(\frac{B_{\star}}{\rm 100\, kG}\right)^{\frac{4}{7}},
\ee

\noindent
where we take $R_{\rm WD} \approx 5.4\times 10^{8}$ cm for the WD radius and normalize to an accretion rate $\dot{M}$ characteristic of CVs \citep{Patterson84}. Depending on the values of $\dot{M}$ and $B_{\star} \sim 10^{5}-10^{7}$ G (in the range of intermediate polars), we typically have $R_{\rm out}/R_{\rm in} \sim 1-10$.

\section{Outburst Photometry}\label{sec:phot}
\subsection{ASAS-SN Pre-Discovery Detections} \label{sec:asas-sn}

The All-Sky Automated Survey for Supernovae (ASAS-SN; \citealt{Shappee+14, Kochanek+17}) detected the brightening of \nova\ prior to its discovery. We use the online ASAS-SN light curve generation interface\footnote{https://asas-sn.osu.edu} to extract aperture photometry (Table~\ref{tab:asas-sn}). The \texttt{bq} camera, which is located at Cerro Tololo International Observatory (CTIO), obtained three consecutive 90\,s exposures of the field containing \nova\ on June 12. The first exposure at $t_{\rm ref} + 0.1944$\,d provides a $4\sigma$ detection, and the following two exposures provide $\sim$$7\sigma$ detections. None of the prior 425 observations with the \texttt{bq}  camera that have $\rm{FWHM} < 1.9$\,pixels (15 arcsec) yield a detection at the $4\sigma$ level or higher. To test if neighboring sources contribute significantly to the aperture fluxes we compute the weighted average flux of these 425 images and find that it is consistent with zero ($0.000 \pm 0.007$\,mJy). There are several stars within 15 arcseconds of \nova\ that should contribute significantly to the aperture sum; however, the field is crowded and it is likely that the sky sample used to estimate the background level contains a similar distribution of field stars. Thus we conclude that the ASAS-SN flux measurements are representative of \nova\ alone and any light from neighboring sources has been removed through normal background subtraction. 

The ASAS-SN light curve generator has an option to measure the flux in stacked images after image subtraction by a reference template. With image subtraction a magnitude of $g = 17.111 \pm 0.136$ is found, but following a similar analysis as above we also find the prior stacked epochs yield a significantly negative average flux of $-0.289 \pm 0.006$\,mJy. Thus we believe the image subtraction code is over subtracting the local background and we favor the simple, aperture photometry results on the individual images.

\subsection{Evryscope Observations} \label{sec:evryscope}

The early-phase, quick-extraction light curve of \nova\ as recorded by Evryscope North at the Mount Laguna Observatory has previously been reported \citep{quimby2021,sokolovsky2023}. These data were extracted using aperture photometry and they exclude epochs where any single pixel reached the saturation limit. Here we reanalyze these images to greatly improve the photometry using point spread function (PSF) techniques. 

Evryscope North is a system of 20 wide-angle cameras delivering a total field covering 7409 square degrees with 13.1 arcsec pixels down to around 16th magnitude in the $g$-band every two minutes \citep{law14}. Images are processed to remove bias, corrected for flat-field variations, and astrometric solutions are determined using a custom built reduction package \citep{corbett2023}. 

Evryscope North began observing the field of \nova\ as it rose above airmass 1.5 starting at $t_{\rm ref} + 0.252$\,d. This was 11.4 hours before it reached maximum optical brightness and 7.2 hours before the discovery images were recorded by Seiji Ueda of Japan (CBET 4976).
Evryscope North recorded a sequence of sixty exposures, each 2 minutes in duration with $\sim 4$\,s between exposures for readout. Normally the system would rotate to a new sky position at the end of each two hour exposure block and then begin new observations with most fields being imaged by different cameras in the system. However, due to an unknown error the system paused for 50 minutes after the first exposure block before resuming regular operations. In total, three separate cameras recorded the rise of \nova\ on the first night of outburst, and four cameras total observed the field on the days before and after the nova outburst. 

\begin{figure*}
    \centering
    \includegraphics[width=2\columnwidth]{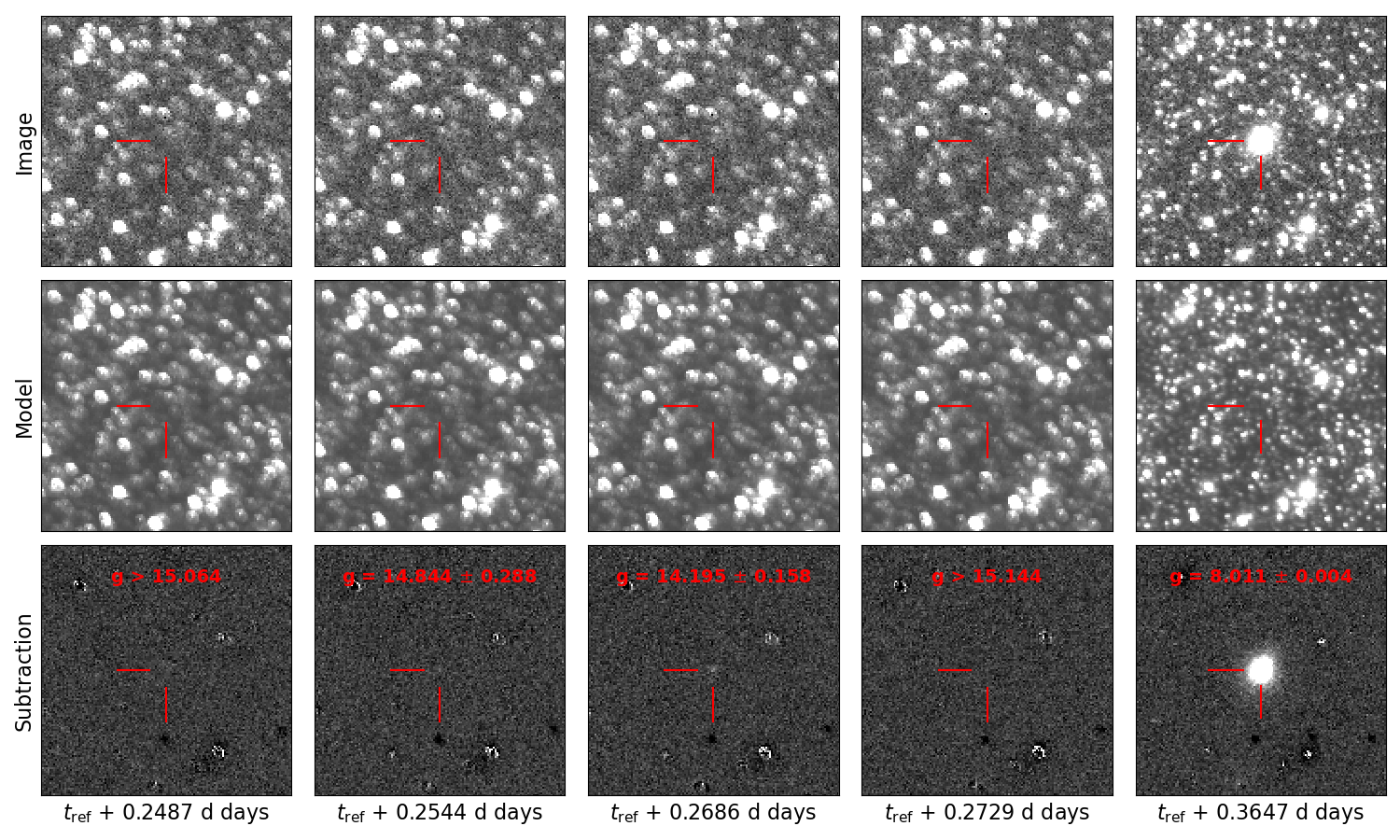}
    \caption{Images centered on V1674 (top row), reference image model (middle row), and corresponding subtractions (bottom row) for select Evryscope observations on June 12th (UT). From left to right: first Evryscope limit ($t_{\rm EVR, 0}$); first Evryscope detection ($t_{\rm EVR, 0}$ + 8.2 min.); early peak ($t_{\rm EVR, 0}$ + 28.7 min.);  the dip ($t_{\rm EVR, 0}$ + 34.9 min.); first image after the Evryscope ratchet ($t_{\rm EVR, 0}$ + 167.1 min.). Each image is $30^{\prime} \times 30^{\prime}$ and has North up and East left. The red cross hairs mark the position of the nova.}
    \label{fig:subtractions}
\end{figure*}

\nova\ is located in a crowded field ($b \sim 6^{\circ}$), and it first appeared in a section of the Evryscope field where the PSF is complex, extended, under-sampled, and highly spatially variable (see Fig.~\ref{fig:subtractions}). To extract accurate photometry from these data, we have developed new methods to model the PSF and to remove light from contaminating sources. A full description of this procedure is planed for a future paper. In \S~\ref{sec:imsub} and \S\ref{sec:evr_phot} we briefly describe the method and present testing to verify the accuracy of the photometry we extract for \nova.

\subsubsection{Image Model Subtraction} \label{sec:imsub}

In classic image subtraction, a reference image (REF) is spatially aligned with a new science frame (NEW) and then convolved with a kernel so that the resulting, convolved REF has the same effective PSF as the NEW. The aligned, convolved REF can then be subtracted from the NEW to produce a residual frame, SUB, that ideally conveys how objects have brightened or faded between the REF and NEW epochs. In reality, SUB frames are typically pocked by errors in the alignment and convolution process that greatly outnumber real astrophysical changes on the sky. These defects are at their worst when the PSF is undersampled and spatially variable, as is the case for the Evryscope North data. 

To address this, we separate the PSF into a component that varies with position, $P_{xy}$, and a component that changes with each exposure, $P_t$. The effective PSF including the pixel response, $P_{xyt}$, is the convolution of these two,
\begin{equation}
P_{xyt} = P_{xy} \ast P_t
\end{equation}
For Evryscope, the coarse ($13.1^{\prime\prime}$) pixel scale means that $P_t$ is not significantly dependent on atmospheric seeing 
and is dominated by small tracking errors. We fit iteratively for $P_{xy}$ and $P_t$ by dividing the CCD into small cells over which we assume $P_{xy}$ is constant and then finding the single $P_t$ for each exposure that best transforms the $P_{xy}$ in each cell to match the data in each cell. With this approach we can assume the PSF to be constant over a much smaller area than would be needed to obtain a set of stars with sub-pixel dither. We only need one star per cell, as we have hundreds of images of each cell. Evryscope tracks the sky for 2 hour blocks, and small tracking errors due to imperfect polar alignment ensure that stars will be dithered as required to properly sample $P_{xy}$, and we employ enough cells to properly sample $P_{t}$ on each image.

After we have modeled $P_{xyt}$ over the field, we use an external photometric catalog, ATLAS Refcat2 \citep{tonry2018}, to set the locations and relative fluxes of sources within each Evryscope field, and then, using the $P_{xyt}$ model, generate a model reference image which can be used to remove quiescent light from the Evryscope images. We refer to this process as, ``image model subtraction''. Examples of this technique are shown in Figure~\ref{fig:subtractions}. The limiting magnitude of Refcat2 ($g \lesssim 19$) is significantly deeper than the Evryscope images ($g \lesssim 16$), which ensures that even sources with marginal flux contributions are included in the reference model. Extended sources are not included in the reference model; however, no such sources are noted in the vicinity of \nova. There are slight effective filter differences between the Everyscope $g$-band and the Refcat2 $g$-band which are factored into the model by assuming:
\begin{equation}
   g_{\rm evr} =  g_{\rm ps} + C_1 X + C_2 (g_{\rm ps} - r_{\rm ps})
\end{equation}
where $g_{\rm evr}$ is the Evryscope $g$-band magnitude, $g_{\rm ps}$ and $r_{\rm ps}$ are the $g$ and $r$ band magnitudes from Refcat2, respectively, $X$ is the airmass, and $C_1$ and $C_2$ are constants determined from a least-squares fit.

\subsubsection{Photometry} \label{sec:evr_phot}

We produce PSF-fit photometry by scaling $P_{xyt}$ to best match the target flux on each residual image in the least-squares sense (the target is excluded from the image model, so the measured flux represents the actual flux of the target and not the difference from some other epoch). The uncertainty in this scaling is used to set the flux uncertainty. The resulting flux measurements and $g$-band magnitudes in the ATLAS Refcat2 system \citep{tonry2018} are provided in Table~\ref{tab:photometry}. We exclude bad or saturated pixels from the PSF fitting, which may increase the systematic uncertainty especially for $g < 10$\,mag stars (Table~\ref{tab:photometry} includes only statistical uncertainties). Note that Evryscope has anti-blooming detectors, so when a pixel saturates it does not affect its neighbors.

The photometry originally presented in \citealt{quimby2021} and \citet{sokolovsky2023} bears some qualitative resemblance to our revised analysis, but there are important differences. First, the initial $\sim 40$ minutes of observations is now better seen as a slow rise instead of the plateau first reported. The magnitude of these observations are also significantly ($\sim 1$ mag) fainter than previously reported. Evidence for a dip in brightness before the main outburst remains, and a clear peak is now detected prior to this dip. These differences are likely due in part to contaminating sources on the original, aperture photometry pipeline.

An advantage of image model subtraction is that we have complete control over which sources are included on the reference model and which are not. Thus to validate our photometry we can select any star in the field, exclude it from the reference model, and then measure the residual brightness on the subtracted image. We perform this procedure for a few thousand stars in the field and then use the measured residuals to determine the 50\% completeness limit for each image.
These limits are provided along with the photometry in Table~\ref{tab:photometry}. Additionally, we produce light curves for these field stars and use them to check the error estimates by comparing the average error to the standard deviation of the light curve over a broad range of mean source brightness. We find that the error bars for $g > 14$\,mag stars are typically overestimated by about 30\%. By $g \sim 12$\,mag the measured uncertainties are well matched to the measured light curve standard deviations, and brighter stars have underestimated uncertainties.

\section{Light Curve Analysis} \label{sec:lc}

The rising portion of the \nova\ outburst captured on June 12th by ASAS-SN and Evryscope can be divided into at least three parts: 1) an initial $\sim$5\,mag brightening compared to the quiescent ZTF measurements in 2021; 2) a subsequent, one hour period over which the flux increases by a factor of $\sim$11.5 in a roughly linear fashion; and 3) the next 3.5 hours over which the flux continues to increase linearly with time but at a higher rate. We refer to these phases, respectively, as the ``slow rise,'' ``fast rise'', and ``faster rise,''  and discuss each of these in the following sections.

\subsection{Slow Rise}
The first pre-discovery detections by ASAS-SN and Evryscope demonstrate a flux increase of about 5.5 times over 1.8 hours. The first ASAS-SN measurement is significantly fainter than the second, which indicates variability on minute timescales. The 105\,s spacing between the first and second ASASS-SN obsevations corresponds to roughly 1/5 of the spin period measured by ZTF. If the spin period is constant from the pre-outburst ZTF measurements, the midpoints of these two exposures should not represent extreme phases of the prior spin variability (see Fig.~\ref{fig:spin_period}). However, the relative variability in the ASAS-SN photometry is quite similar to the historical ZTF trend if we arbitrarily shift the phase of the ASAS-SN measurements by 1/5 of the spin period. Thus, it is possible that the variations in the ASAS-SN light curve are related to the spin of the WD with a difference in phase compared to the ZTF light curve.

The first four Evryscope North observations yield low-significance detections (formally $\sim2.3\sigma$, but correcting for the over-estimated uncertainty noted in \S\ref{sec:phot} the significance would be above $3\sigma$), and the target is seen to brighten significantly at an average rate of $0.108 \pm 0.021~\rm{mJy}~\rm{min}^{-1}$ through the 14th measurement. The following observation with a mid-exposure time of $t_{\rm ref} + 0.2726$\,d
is roughly $2\sigma$ brighter than this trend and is followed by a fading at a rate of $1.030 \pm 0.052~\rm{mJy}~\rm{min}^{-1}$ over the next 6.1 minutes, which culminates in the lowest significance detection (formally $\sim1.1\sigma$) at 
$t_{\rm ref} + 0.2769$\,d.
We refer to this epoch as the ``dip'' and to the prior maximum as the ``early peak'' (Fig.~\ref{fig:early_lc}). 

Prior to the early peak the flux increase shown by Evryscope is consistent with a linear rise. However, extrapolating this trend back in time would predict negative flux at the ASAS-SN epochs. A joint fit to the ASAS-SN and first 14 Evryscope measurements results in an assumed linear brightening rate of $0.0324 \pm 0.0015~\rm{mJy}~\rm{min}^{-1}$, which is significantly slower than the rate derived from the first 14 Everyscope points alone. If we assume the initial flux increase can be fit with a power law:
\begin{equation}\label{eqn:powerlaw}
    f(t) \propto (t - t')^n
\end{equation} we find that larger values of $n$ are preferred, but do not significantly improve the match to the steeper Everyscope rise as compared to $n=1$. 

\subsection{Fast Rise}
After the dip a dramatic brightening phase begins at
$t_{\rm ref} + 0.2812$\,d
with a flux increase factor of $2.3 \pm 0.6$ over the previous Evryscope exposure. The brightening rate rapidly approaches an average of $13.80 \pm 0.07$\,mJy\,min$^{-1}$ and remains nearly constant for about 1 hour. Over this period the flux increases by a factor of 185 with typical deviations from the linear model of only 1\%. We fit Equation~\ref{eqn:powerlaw} to these data and find a best-fit power-law index of $n = 1.07 \pm 0.01$ (and $t' = t_{\rm ref} + 0.2824 \pm 0.0002$\,d),
which indicates a nearly linear rise is favored. For comparison, the initial brightening rate of T Pyx, which is known for its rapid initial rise \citep{schaefer2013}, averages to just $\sim$1.0\,mJy\,min$^{-1}$ for a factor of 2 increase over a 1.8\,hour period. 

\subsection{Faster Rise}
The first two-hour observing block ends with 4 measurements that deviate slightly from the fast rise trend. When the next camera began observing again after the 50\,min technical delay the measured fluxes are $\sim$30\% brighter than predicted by the fast rise trend. This indicates a break in the light curve to a higher rate of flux increase occurs near the end of the first observing block. 
The rate of brightening is about 63\% higher compared to the fast rise phase. Fitting only measurements between 
$0.3282\,{\rm d} < t - t_{\rm ref} < 0.4591$\,d
we find a best-fit power-law index of $0.97 \pm 0.02$ (and $t' = t_{\rm ref} + 0.3020$\,d),
which is again consistent with a linear rise, this time at a rate of $22.46 \pm 0.11$\,mJy\,min$^{-1}$. We cannot rule out a small ($\sim$0.05\,mag level) systematic offset between the first and second camera measurements, but such a shift would have minimal impact on the linearity of the rise. We also note that a third Evryscope camera began observing the field of \nova\ starting at 
$t_{\rm ref} + 0.4535$\,d.
We truncate the data from this camera at 18 degree morning twilight, but the 5 observations prior to twilight appear consistent with the same linear trend. 

Extrapolating this linear rise, we would expect \nova\ to be near $g=6.7$\,mag at the time of discovery ($t_{\rm ref} + 0.5411$\,d); however, the reported discovery magnitudes are 8.4 and 8.0 on the discoverer's Canon EOS 6D digital camera. We find it likely that the nova is saturated on the discovery images because shortly after they were taken, the nova was confirmed by Koichi Itagaki at a CCD magnitude of 6.7 at 
$t_{\rm ref} + 0.5551$\,d.
This is about 0.1\,mag fainter than a continuation of the linear brightening observed by Evryscope predicts, which is likely within systematic uncertainties. Thus, \nova\ may have continued rising roughly linearly in the optical through Mr. Itagaki's confirmation image.

\section{Nova Models} \label{sec:models}

In this section we discuss models that could potentially explain features of the rising light curve. We consider heating of the WD photosphere by thermonuclear fusion in \S\ref{sec:tnr}. In \S\ref{sec:toy_model} we develop a toy model based on expansion of the WD envelope at constant luminosity and use it to place physical constraints on the outburst. In \S\ref{sec:reprocessing} we consider how X-rays emitted from the hot WD envelope may be reprocessed into optical light by the accretion disk or companion, and we introduce a three-phase toy model in \S\ref{sec:toy_model3} with constant velocities in each phase to represent slow, fast, and faster expansion of the WD envelope.

\subsection{Thermonuclear Heating}\label{sec:tnr}
In the classic picture of a nova, when the convective envelope reaches the WD's photosphere its temperature begins to increase but at first the radius remains roughly constant (e.g., \citealt{Prialnik+79}). The WD only begins to expand when the bolometric luminosity reaches the Eddington limit, which is about $1.4 \times 10^{38}$\,erg\,s$^{-1}$ for a solar mass WD. Assuming a WD radius of $R_0 = 5400$\,km, this luminosity is reached when the effective surface temperature reaches a peak of about $T_{\rm max} = 9\times10^{5}$\,K. Thus before expansion sets in the absolute $g$-band magnitude of the WD would be about $M_g \sim 6.3$. For the distance and line-of-sight absorption appropriate to \nova\ ($D \gtrsim 3$\,kpc; $A_g=1.9$\,mag), the predicted brightness at $T_{\rm max}$ is $g \gtrsim 20.6$\,mag. This limit is comparable to observations of the system in quiescence and, critically, much too faint to explain the early ASAS-SN and Evryscope observations. We next consider if the early detections are consistent with expansion of the WD.

\subsection{Toy Model: Expansion at Constant Luminosity}\label{sec:toy_model}
As thermonuclear fusion drives the luminosity of the nova to the Eddington limit the WD begins to expand, but the bolometric luminosity settles at a roughly constant value (e.g., \citealt{Gallagher&Code74}). Thus the WD surface temperature cools from $T_{\rm max}$, and the peak of the SED moves from the X-ray to the UV and finally into the optical. We now construct a toy model to represent this process and compare it 
to the observations. 

For a blackbody expanding at constant luminosity the temperature, $T$, will decrease as the radius, $R$ increases following\begin{equation}\label{eqn:toy_temp}
    T = T_{\rm max} \left(R_0 \over R\right)^{1/2}
\end{equation} 
Assuming the expansion velocity quickly reaches a steady velocity,
$v$, we can write\begin{equation}\label{eqn:toy_rad}
    R = R_0 + v(t - t_0'') ,
\end{equation} for $t > t_0''$, where $t_0''$ is the time at which the expansion begins. The toy model then predicts an observed flux density given by
\begin{equation}
    f_{\nu} = \pi B_{\nu, T} \left(R \over D\right)^2 ,
\end{equation} where $B_{\nu, T}$ is the Planck function for temperature $T$. For any given time we can then set the radius and temperature using equations \ref{eqn:toy_temp} and \ref{eqn:toy_rad}, compute $f_\nu$, and then convolve this spectral energy distribution with an optical band pass to recover the classic light curve of a nova outburst characterized by a smooth rise to peak followed by a slower decline.

\begin{figure*}
    \centering
    \includegraphics[width=2\columnwidth]{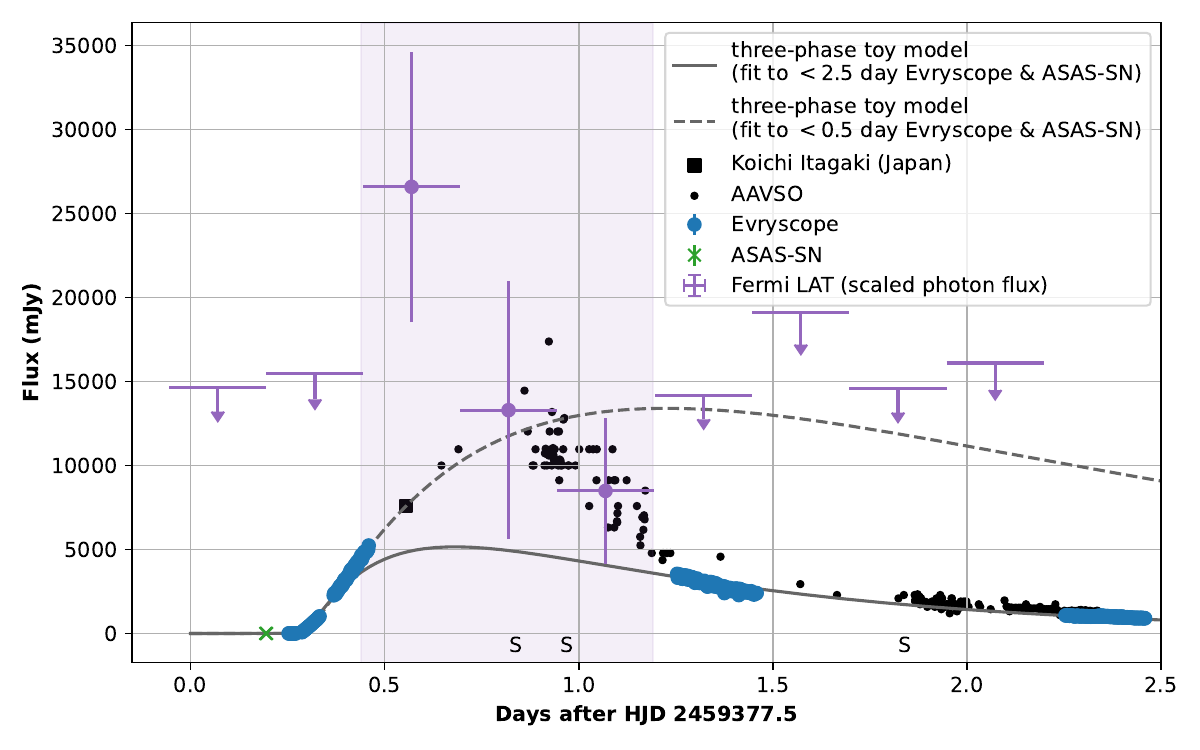}
    \caption{Three-phase toy model fits (grey curves) to the ASAS-SN and Evryscope measurements. Black points are AAVSO measurements \citep{aavso}, either visual or in the V-band, which are plotted as equivalent to $g$-band fluxes assuming no systematic offsets and are not included in the fitting. Purple dots give the Fermi/LAT photon flux from \citet{sokolovsky2023} arbitrarily scaled to align the last two detections with the AAVSO measurements. The time span covering the Fermi/LAT detections is highlighted, and limits are shown outside this range with down pointing arrows. The first reported epochs of optical spectroscopy are marked with ``S'' \citep{mun21, Aydi+21}.}
    \label{fig:toy_model}
\end{figure*}

However, this toy model cannot even qualitatively explain the distinct sections recorded in the rise of \nova. The toy model lacks the initial, slow rise and transition into a fast rise exhibited by \nova, and it further lacks the second transition to a faster rise before the peak. The slow decline of the toy model relative to its fast rise also does not appear to match the rather symmetric rise and decline of \nova\ around peak (see Fig.~\ref{fig:toy_model}). 

We next consider the brightness predicted by the toy model at key stages of the nova's development. The observed flux may exceed these predictions if other sources of power are present, but we can use the toy model to rule out some situations where the observed fluxes fall below the toy model predictions.

We begin by considering the expected brightness of the WD as it expands to fill its Roche lobe. With the bolometric luminosity fixed at the Eddington limit and approximating the emitting surface using the volume equivalent radius \citep{eggleton1983},
the $g$-band magnitude would be between about 14.9\,mag and 13.4\,mag for distances of 6\,kpc to 3\,kpc, respectively, assuming $A_g=1.9$. Thus at the higher distance the predicted flux roughly corresponds to the brightness first reached near the early peak. To put it another way, the WD was unlikely to have filled its Roche lobe prior to the early peak. 

Another important radius to consider is the outer radius of the accretion disk, $R_{\rm out} \approx 1.9\times 10^{10}\,{\rm cm},$ (see \S\ref{sec:physical}). Again considering distances of 3\,kpc to 6\,kpc, the toy model predicts $14.9\,{\rm mag} < g < 16.4\,{\rm mag}$. Thus, the observed slow rise to peak could be consistent with the WD expanding to $R_{\rm out}$ if the distance to the nova is at the low end. In this case, the implied expansion velocity averaged just 20\,km\,s$^{-1}$ over the 2.3\,hours before the fast rise. This is orders of magnitude slower than the escape speed from a $\sim$1.0\,$M_{\odot}$ WD at a radius of $R_{\rm out}$.

When the temperature cools to roughly $1.5 \times 10^5$\,K, the opacity increases due to the presence of iron-peak elements, and this is expected to launch a fast, line-driven wind \citep{kato_hachisu1994}. With our assumed $R_0$ and $T_{\rm max}$, the toy model predicts that this temperature is reached at roughly $R_{\rm out}$. Thus it is plausible that the fast rise following the initial slow rise in the Evryscope light curve represents the launch of the fast wind.

\subsection{Reprocessing by Accretion Disk and Secondary}\label{sec:reprocessing}

Another important process to consider at the earliest phases of the nova outburst is the reprocessing of X-rays emitted by the hot WD in the accretion disk or irradiated side of the companion.
We start by considering reprocessing by the accretion disk.  At a given distance $r$ from the WD, a fraction $\sim H/R_{\rm out}$ of the total solid angle is subtended by the accretion disk, where $H$ is the vertical scale-height of the disk and the aspect ratio $H/r$ is calculated near the outermost radius of the disk $r \approx R_{\rm out}$ because $H/r$ typically increases with $r$.  The reprocessed luminosity is therefore given by
\be
L_{\rm rep} \sim (H/R_{\rm out})L_{\rm WD}(1-\mathcal{A}),
\ee
where $\mathcal{A}$ is the albedo of the disk surface, which we expect to be of order unity because of the significant neutral fraction of the disk material (e.g., \citealt{vanparadijs83}).  Prior to the nova outburst, we expect $H/r \sim 0.03$ on radial scales $r \sim R_{\rm out}$, depending weakly on the disk viscosity parameter $\alpha$ and $\dot{M}$ (e.g., \citealt{Frank+02}; their Eq. 5.49).  However, $H/r$ may be somewhat larger than this estimate as a result of irradiation heating dominating over internal viscous heating of the disk.

Equating the flux of radiation $F_{\rm heat} \simeq L_{\rm rep}/(2\pi r^{2})$ to the rate of radiative cooling $F_{\rm cool} \simeq 2\pi \sigma T_{\rm rep}^{4}$, gives an estimate of the effective temperature for the reprocessed disk emission,
\begin{eqnarray}
T_{\rm rep} &\simeq& \left(\frac{(H/r)L_{\rm WD}(1-\mathcal{A})}{4\pi^{2}\sigma r^{2}}\right)^{1/4}
\approx 3.9\times 10^{4}\,{\rm K}\,\times \nonumber \\
&& \left(\frac{R_{\rm out}}{2\times 10^{10}\,{\rm cm}}\right)^{-1/2}\left(\frac{L_{\rm WD}}{1.4\times 10^{38}\,{\rm erg\,s^{-1}}}\right)^{1/4} \times \nonumber \\
&& \left(\frac{H/r}{0.03}\right)^{1/4},
\end{eqnarray}
where we have taken $\mathcal{A} = 0.5$ (e.g., \citealt{vanparadijs83}) and normalize $r$ to $R_{\rm out}$ from Eq.~\ref{eq:Rout}.

Blackbody emission of temperature $T_{\rm rep} \approx 4\times 10^{4}$ K peaks above the optical $g$-band $\nu_{\rm opt} \approx 6\times 10^{14}$ Hz of our observations.  If both the WD and the reprocessed emission are observed in an optical band on the Rayleigh-Jeans tail of a blackbody (i.e., $h\nu \ll kT_{\rm rep}, kT_{\rm WD}$), the ratio of their optical band luminosities can be written,
\be
\frac{L_{\rm opt,rep}}{L_{\rm opt, WD}} \approx \frac{R_{\rm out}^{2}T_{\rm rep}}{R_{\rm WD}^{2}T_{\rm WD}} \approx 13.5\left(\frac{H/r}{0.03}\right)^{1/4}\left(\frac{R_{\rm WD}}{0.03R_{\rm out}}\right)^{-3/2}
\ee
The reprocessed luminosity can thus dominate the optical light curve, until the white dwarf has begun to expand significantly from its initial radius ($R_{\rm WD} \gtrsim 0.2R_{\rm out}$).  After this point, the white dwarf photosphere will soon come to dominate and the light curve will begin to rise faster. This scenario could possibly be at play in the slow-rise phase with the ASAS-SN measurements dominated by reprocessed light and the Evryscope observations dominated by the WD photosphere. Alternatively, if the WD engulfs the disk somewhat prior to dominating the optical luminosity this might create a brief dip in the light curve after the disk is destroyed, similar to what is observed.  

The companion star can be an equally if not more important source of reprocessing (e.g., \citealt{vanparadijs&mcclintock94}).  In particular, the ratio of the solid angle subtended by the secondary to that subtended by the disk is given by
\be
\frac{\Omega_{\star}}{\Omega_{\rm d}} \approx \frac{\pi R_{2}^{2}/a_{\rm bin}}{2\pi R_{\rm out}(H/r)} \approx 4\left(\frac{R_2}{0.3R_{\odot}}\right)^{2}\left(\frac{H/r}{0.03}\right)^{-1},
\ee
where $R_2 \approx 0.3R_{\odot}$ is the characteristic radius of CV donors of mass $M_2 \approx 0.3M_{\odot}$ (e.g., \citealt{Knigge06}) and we have taken $a_{\rm bin} \approx 8\times 10^{10}$ cm for the binary separation.  Reprocessing by the donor star may even dominate that by the accretion disk, depending on $H/r$.  Indeed, periodic modulation of nova light curves due to irradiation of the companion, has been reported for V1974Cyg \citep{DeYoung&Schmidt94}, V407Lup \citep{Aydi+21}, V392 Per \citep{Munari+20,Murphy-Glaysher+22}; however, in most cases this modulation is observed {\it after} the peak of the outburst, 2-4 mag below the peak brightness (see \citealt{Schaefer22b} for a review).

\subsection{Three-Phase Toy Model}\label{sec:toy_model3}

Motivated by the three distinct parts of the observed rise to peak we next consider a version of the toy model in which the expansion rate has 3 parts. We adopt the same assumptions as before, namely that the nova is a blackbody emitter and that the temperature drops as the radius increases to maintain a constant bolometric luminosity. We maintain the simplification that the photospheric radius increases linearly with time, but here we allow the photospheric velocity in each phase to differ. We can then fit for the times each phase begins and the velocities of the photosphere in each phase, and optionally the distance, initial radius, $R_0$, and initial temperature, $T_{\rm max}$ at the start of the expansion phase.

\begin{figure}
    \centering
    \includegraphics[width=\columnwidth]{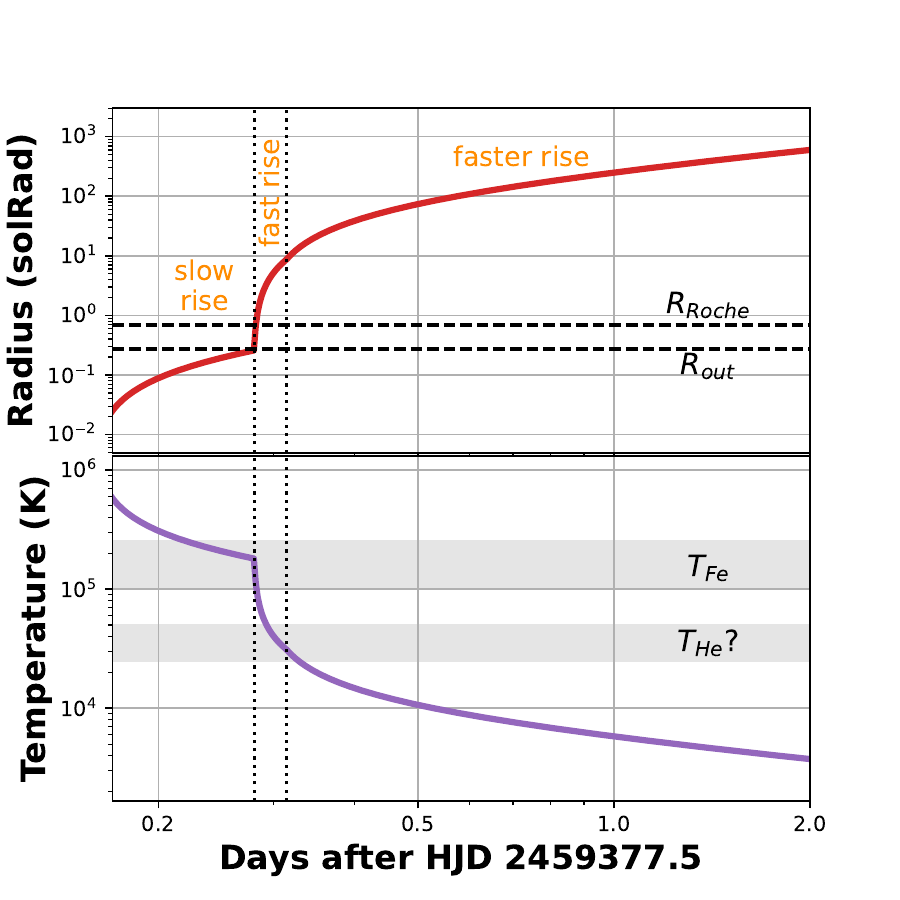}
    \caption{Radius and temperature evoloution of the best-fit three-phase toy model. Dashed lines mark the estimated, volume-equivalent Roche lobe radius, $R_{\rm Roche}$, and the outer disk radius, $R_{\rm out}$. Approximate temperature ranges where higher opacities are possible (depending on density) are indicated by the light-gray bands.
    }
    \label{fig:rad_temp}
\end{figure}

The best-fit three-phase toy model is shown in Figures~\ref{fig:early_lc} and \ref{fig:toy_model}, and the best-fit radius and temperature evolution is shown in Figure~\ref{fig:rad_temp}. The fit favors a low distance, $D=2.8$\,kpc, but the precise value is uncertain given that distance is degenerate with luminosity (and thus $R_0$ and $T_{\rm max}$). It is suggestive, however, that the best-fit temperature curve drops to near the range where the opacity is expected to increase significantly due to iron-peak elements (about $1.5 \times 10^5$\,K) at roughly the same time that the fast rise begins. The best-fit velocity for the fast rise is about 2000\,km\,s$^{-1}$, and this increases to about 2800\,km\,s$^{-1}$ in the faster rise phase, which is reminiscent of the 3000\,km\,s$^{-1}$ absorption features observed in the early spectra \citep{mun21, Aydi+21}.

Although the three-phase toy model can account for the slow and fast rise phases and it can account for the observed light curve in the first days following peak, it is not able to fit \nova\ around the time of optical maximum and it begins to fade faster than the data starting a couple days after peak. Observations of $\gamma$-rays and X-rays over these periods have implied the presence of shocks, which may further bolster the optical flux from \nova\ over the predictions of the three-phase toy model. In Figure~\ref{fig:toy_model} we also show a three-phase toy model fit to the earliest, $t < t_{\rm ref} +0.5$\,d data only. In this case the post-peak brightness is greatly over predicted, and the best-fit distance, 1.3\,kpc, is much lower.

\section{Discussion and Conclusions} \label{sec:conclusions}

The first ASAS-SN detections combined with the high-cadence monitoring of Evryscope provide an unprecedented picture of the optical rise of a fast nova. The departure from quiescence may have begun with the anomalously bright detection of \nova\ by ZTF on $t_{\rm ref} - 1.5746$\,d, which is 2.4\,d before the estimated date of $V$-band maximum (Appendix~\ref{sec:t2}). The following observations by ASAS-SN starting at $t_{\rm ref} + 0.1944$\,d mark a $\sim$3\,mag rise in brightness over the ZTF baseline and show variability on timescales consistent with the WD spin period but out of phase. \nova\ was observed with a 2\,min cadence by Evryscope starting at $t_{\rm ref} + 0.2527$\,d when it was already $\sim$1\,mag brighter than the ASAS-SN average. It reached an early peak minutes later, and then faded significantly before rebounding and then starting a fast, almost linear rise in flux for 1 hour. This was followed by a second break to a faster, even more linear rise through at least the next 3 hours when morning twilight began for Evryscope.

The rough consistency between the initial ASAS-SN detections and the shape of the historical ZTF phase-folded light curve implies that the accretion spot remains a dominant source of power at these early phases of the outburst. If the $g$-band flux scales linearly with the accretion rate, $\dot{M}$, then $\dot{M}$ increases by a factor of $\sim$15 over the historical ZTF rate. Such a scenario might occur during the X-ray flash as the X-ray luminosity of the WD increases and begins ablating the companion, similar to what \citet{Patterson+22} argue takes place during the soft X-ray phase after optical maximum. Photo-evaporation of the disk might also lead to an increase in the aspect ratio of the disk, $H/r$, which would increase the reprocessed luminosity and perhaps explain the brightening trend from the first ASAS-SN detections to the first Evryscope detections. Alternatively, these ASAS-SN and Evryscope observations are separated by roughly half the orbital period, so the change in brightness might reflect the projected area of the reprocessing region, although some fine-tuning is required in this case to have the peak immediately prior to the start of the fast rise.

Assuming instead that the rise of \nova\ is powered exclusively by changes in the temperature and radius of the WD photosphere, we can parameterize the slow, fast, and faster rising phases with a three-phase toy model. The model assumes a blackbody radiating constantly at the Eddington luminosity and thus cooling as the photosphere expands slowly at first, then fast, then faster. The best-fit model begins with a gentle expansion characterized by a photospheric velocity, $v_{\rm ph,1}$, of $\sim 20$\,km\,s$^{-1}$, and then transitions to a fast wind, $v_{\rm ph,2}\sim 2000$\,km\,s$^{-1}$. At the transition point, the model favors a photospheric temperature of around $T_{\rm ph}\sim 1.5 \times 10^{5}$\,K, which is in the range where iron-peak elements produce an opacity peak (see Fig.~\ref{fig:rad_temp}). Such an increase in opacity has been discussed as the impetus for an optically thick wind \citep{kato_hachisu1994}. The best-fit toy model has a photospheric radius of $\sim0.26$\,$R_{\odot}$ at the point where the photospheric velocity increases to $v_{\rm ph,2}$. The escape velocity for a $1.3$\,$M_{\odot}$ WD with this radius is $v_{\rm esc} \sim 1400$\,km\,s$^{-1}$, so the toy model provides a roughly consistent picture of a WD envelope cooling into the iron opacity bump and launching an optically thick wind. 

However, the toy model does not predict a dip in flux just prior to the fast rise as is observed in the Evryscope light curve. If the fast rise represents the launch of an optically thick wind brought on by a sudden increase in opacity, then perhaps this sudden opacity increase also momentarily diminishes the $g$-band flux? It is not obvious that this will happen because the photospheric radius will increase and the temperature will further decrease as the wind is launched, which will ultimately increase the optical flux. Models implementing radiation hydrodynamics are needed to study this transition in more detail. If the dip is connected to a sudden increase in opacity, then single-band, high-cadence observations like these from Evryscope provide precise times at which the WD photosphere reaches a known temperature and will thus provide powerful modeling constraints. Alternatively, this transition may occur coincidentally to the envelopment of the accretion disk by the expanding WD envelope, which would quickly put an end to any X-ray reprocessing that may have been supplementing the optical flux, resulting in a $g$-band dip. In this case, observations of other novae may not have a dip immediately prior to the launch of a fast wind.

The third, faster rising portion of the early \nova\ light curve is also difficult to explain in the context of the three-phase toy model. Fitting to only the ASAS-SN and Evryscope data obtained prior to optical maximum, the model can account for the linear rising trends in the Evryscope data and make reasonable predictions for the pre-maximum observations recorded by amateurs; however, it cannot explain the rapid post maximum fading. Physically, we would expect the toy model to break down as the outer envelope thins and the photosphere begins to recede. For comparison, the fastest evolving model considered by \citet{ksh2024}, a 1.3\,M$_{\odot}$ WD with a $2\times 10^{-6}$\,M$_{\odot}$ carbon enriched envelope, takes 2.5 days to reach this point after the shell flash (defined as the moment of maximum nuclear luminosity), whereas the optical peak of \nova\ occurs just 0.64\,days after the first ASAS-SN detection. Thus, if the optical peak corresponds to the time the photosphere is its largest, then the envelope mass is likely significantly smaller than $2\times 10^{-6}$\,M$_{\odot}$. 
Alternatively, perhaps the final, anomalously bright ZTF measurement prior to outburst (see Fig.~\ref{fig:spin_period}) better reflects the start of the shell flash. In this case the observed rise to optical maximum, 2.4\,d, would be an excellent match to the \citet{ksh2024} model.

If we extend the three-phase toy model fitting to include the first 3 days of Evryscope observations, the best-fit model provides an acceptable fit to the data \textit{except} for the period over which $\gamma$-rays were detected by Fermi. This could indicate that the optical peak was powered by the same shocks responsible for the $\gamma$-rays. Previous studies have shown correlations between $\gamma$-rays and optical light around peak \citep{Li2017}, and
\citet{sokolovsky2023} find that the ratio of $\gamma$-ray to optical luminosity extrapolated over the optical peak of \nova, $L_{\gamma} / L_{\rm opt} \sim 1 / 500$, is consistent with other $\gamma$-ray novae. \citet{munrai2017} have shown that some light curves can be decomposed into fireball and $\gamma$-ray components. \nova\ may be a similar event. The transition from a fireball dominated light curve to a $\gamma$-ray dominated light curve may also explain the pre-maximum halts observed 0.5--2\,mag below peak in some \citep{hou10} but not all  (e.g., \citealt{thompson2017}) novae. These halts have alternatively been explained as a momentary dip in flux resulting from the convective zone dropping below the WD's photosphere as mass-loss begins \citep{hil14}. The Evryscope light curve of \nova\ rules out such a pre-max halt within 1--8\,mag of peak, but the same physical process might explain the early dip even though it is an unprecedented 8.5\,mag below peak brightness.

The detection of $\gamma$-rays can be understood in connection to the photospheric velocity increase implied by the best-fit three-phase toy model if shocks are produced from interaction of the second, faster wind with the first, slower wind. The velocity of an optically thick wind is typically connected to the escape velocity. As the photosphere expands, this velocity decreases. In the models of \citet{hachisu_kato2022}, shocks are formed after optical maximum as the photosphere begins to recede and the wind velocity thus increases and begins shocking the prior wind. Thus they predict the $\gamma$-rays should begin after optical maximum, but for \nova\ the first Fermi detection comes prior to optical maximum. However, \citet{quataert2016} have discussed internal shocks in outflows driven by (locally) super-Eddington winds. In this scenario the initial wind essentially clears a path for the later wind, which can then run through pre-expanded, lower-density material at higher velocities. Shocks are then produced when this late wind interacts with the early wind, which may explain the early $\gamma$-ray emission in \nova.

The $\gamma$-ray emission from previous novae has been explained in the context of a fast wind interacting with a slow torus of gas that is formed when the WD overflows its Roche lobe and material is ejected by the companion. If this is the case for \nova, then the linear rise in flux constrains the density profile of the slow torus to be $\rho \propto r^{-3}$, which is close to radial behavior expected close to the WD \citep{pejcha2016}. In this case the faster wind would need to be launched by $t_{\rm ref} + 0.314$\,d, which is only 1.9\,d after the anomalously bright ZTF measurement. However, it is difficult to explain the Evryscope light curve in the context of the fast-wind, slow-torus paradigm because the ASAS-SN and Evryscope observations are fainter than the toy model predicts the WD to be during RLOF until the fast rise begins, and this transition comes only hours before the first Fermi/LAT detection. Thus, there does not appear to be sufficient time for material to stream out of the L2 nozzle and form the slow torus. 
The spectra of some other novae indicate the presence of such relatively slow moving gas \citep{aydi2020b}, but the lowest velocity features noted in the earliest spectra of \nova\ are P-Cygni profiles with absorption troughs blue-shifted by $\sim$3000\,km\,s$^{-1}$ \citep{mun21, Aydi+21}. Spectra taken the following days exhibit complex line profiles that favor multiple velocity flows \citep{woodward21} instead of a single fast wind interacting with a single slow wind. It thus appears that the WD began its rapid expansion prior to extensive stripping by the companion. If RLOF begins with the fast rise and shocks begin with the faster rise, then the companion has only $\sim$22\% of an orbit to spray material out the L2 nozzle. 

The observational signature marking the start of a nova outburst has long been predicted to take the form of a UV/X-ray flash \citep{starrfield1990}. However, such data remain scarce due to the limited sensitivity and sky coverage of X-ray instruments (but see \citealt{konig2022}). \nova\ demonstrates an alternative path to studying the earliest phases of novae using wide-field, high-cadence optical surveys like Evryscope and the planned Argus Array \citep{law2022}. With a depth of $g<19.6$\,mag at 60~s cadence and a simultaneous $\sim$8000 square degree field of view, Argus will be capable of recording the optical brightening during the rise to $T_{\rm max}$ and optical signatures of the X-ray flash of nearby ($<5$\,kpc) novae.

\begin{acknowledgements}
    We acknowledge with thanks the variable star observations from the AAVSO International Database contributed by observers worldwide and used in this research. The ZTF forced-photometry service was funded under the Heising-Simons Foundation grant \#12540303 (PI: Graham).  B.D.M. acknowledges support from NASA through the Astrophysics Theory Program (grant 80NSSC22K0807) and Fermi Guest Investigator program (grant 80NSSC24K0408).  The Flatiron Institute is supported by the Simons Foundation.  K.J.S. acknowledges support by NASA through the Astrophysics Theory Program (80NSSC20K0544).
\end{acknowledgements}

\facilities{AAVSO, ASAS-SN, Evryscope, ZTF}

\bibliography{paper}{}

\begin{thebibliography}{}
\expandafter\ifx\csname natexlab\endcsname\relax\def\natexlab#1{#1}\fi
\providecommand{\url}[1]{\href{#1}{#1}}
\providecommand{\dodoi}[1]{doi:~\href{http://doi.org/#1}{\nolinkurl{#1}}}
\providecommand{\doeprint}[1]{\href{http://ascl.net/#1}{\nolinkurl{http://ascl.net/#1}}}
\providecommand{\doarXiv}[1]{\href{https://arxiv.org/abs/#1}{\nolinkurl{https://arxiv.org/abs/#1}}}

\bibitem[{{Ackermann} {et~al.}(2014){Ackermann}, {Ajello}, {Albert}, {Baldini}, {Ballet}, {Barbiellini}, {Bastieri}, {Bellazzini}, {Bissaldi}, {Blandford}, {Bloom}, {Bottacini}, {Brandt}, {Bregeon}, {Bruel}, {Buehler}, {Buson}, {Caliandro}, {Cameron}, {Caragiulo}, {Caraveo}, {Cavazzuti}, {Charles}, {Chekhtman}, {Cheung}, {Chiang}, {Chiaro}, {Ciprini}, {Claus}, {Cohen-Tanugi}, {Conrad}, {Corbel}, {D'Ammando}, {de Angelis}, {den Hartog}, {de Palma}, {Dermer}, {Desiante}, {Digel}, {Di Venere}, {do Couto e Silva}, {Donato}, {Drell}, {Drlica-Wagner}, {Favuzzi}, {Ferrara}, {Focke}, {Franckowiak}, {Fuhrmann}, {Fukazawa}, {Fusco}, {Gargano}, {Gasparrini}, {Germani}, {Giglietto}, {Giordano}, {Giroletti}, {Glanzman}, {Godfrey}, {Grenier}, {Grove}, {Guiriec}, {Hadasch}, {Harding}, {Hayashida}, {Hays}, {Hewitt}, {Hill}, {Hou}, {Jean}, {Jogler}, {J{\'o}hannesson}, {Johnson}, {Johnson}, {Kerr}, {Kn{\"o}dlseder}, {Kuss}, {Larsson}, {Latronico}, {Lemoine-Goumard}, {Longo}, {Loparco}, {Lott}, {Lovellette}, {Lubrano},
  {Manfreda}, {Martin}, {Massaro}, {Mayer}, {Mazziotta}, {McEnery}, {Michelson}, {Mitthumsiri}, {Mizuno}, {Monzani}, {Morselli}, {Moskalenko}, {Murgia}, {Nemmen}, {Nuss}, {Ohsugi}, {Omodei}, {Orienti}, {Orlando}, {Ormes}, {Paneque}, {Panetta}, {Perkins}, {Pesce-Rollins}, {Piron}, {Pivato}, {Porter}, {Rain{\`o}}, {Rando}, {Razzano}, {Razzaque}, {Reimer}, {Reimer}, {Reposeur}, {Saz Parkinson}, {Schaal}, {Schulz}, {Sgr{\`o}}, {Siskind}, {Spandre}, {Spinelli}, {Stawarz}, {Suson}, {Takahashi}, {Tanaka}, {Thayer}, {Thayer}, {Thompson}, {Tibaldo}, {Tinivella}, {Torres}, {Tosti}, {Troja}, {Uchiyama}, {Vianello}, {Winer}, {Wolff}, {Wood}, {Wood}, {Wood}, {Charbonnel}, {Corbet}, {De Gennaro Aquino}, {Edlin}, {Mason}, {Schwarz}, {Shore}, {Starrfield}, {Teyssier}, \& {Fermi-LAT Collaboration}}]{ackermann2014}
{Ackermann}, M., {Ajello}, M., {Albert}, A., {et~al.} 2014, Science, 345, 554, \dodoi{10.1126/science.1253947}

\bibitem[{{Aydi} {et~al.}(2020{\natexlab{a}}){Aydi}, {Chomiuk}, {Izzo}, {Harvey}, {Leahy-McGregor}, {Strader}, {Buckley}, {Sokolovsky}, {Kawash}, {Kochanek}, {Linford}, {Metzger}, {Mukai}, {Orio}, {Shappee}, {Shishkovsky}, {Steinberg}, {Swihart}, {Sokoloski}, {Walter}, \& {Woudt}}]{aydi2020b}
{Aydi}, E., {Chomiuk}, L., {Izzo}, L., {et~al.} 2020{\natexlab{a}}, \apj, 905, 62, \dodoi{10.3847/1538-4357/abc3bb}

\bibitem[{{Aydi} {et~al.}(2020{\natexlab{b}}){Aydi}, {Sokolovsky}, {Chomiuk}, {Steinberg}, {Li}, {Vurm}, {Metzger}, {Strader}, {Mukai}, {Pejcha}, {Shen}, {Wade}, {Kuschnig}, {Moffat}, {Pablo}, {Pigulski}, {Popowicz}, {Weiss}, {Zwintz}, {Izzo}, {Pollard}, {Handler}, {Ryder}, {Filipovi{\'c}}, {Alsaberi}, {Manojlovi{\'c}}, {Lopes de Oliveira}, {Walter}, {Vallely}, {Buckley}, {Brown}, {Harvey}, {Kawash}, {Kniazev}, {Kochanek}, {Linford}, {Mikolajewska}, {Molaro}, {Orio}, {Page}, {Shappee}, \& {Sokoloski}}]{aydi2020}
{Aydi}, E., {Sokolovsky}, K.~V., {Chomiuk}, L., {et~al.} 2020{\natexlab{b}}, Nature Astronomy, 4, 776, \dodoi{10.1038/s41550-020-1070-y}

\bibitem[{{Aydi} {et~al.}(2021){Aydi}, {Sokolovsky}, {Chomiuk}, {Strader}, {Kawash}, {Page}, {Boussin}, {Ikonnikova}, {Atapin}, {Belinski}, {Burlak}, {Dodin}, {Maslennikova}, {Postnov}, {Potanin}, {Safonov}, {Shatsky}, {Tatarnikov}, {Korotkiy}, {Stanek}, {Kochanek}, \& {Shappee}}]{Aydi+21}
---. 2021, The Astronomer's Telegram, 14710, 1

\bibitem[{{Aydi} {et~al.}(2023){Aydi}, {Chomiuk}, {Miko{\l}ajewska}, {Brink}, {Metzger}, {Strader}, {Buckley}, {Harvey}, {Holoien}, {Izzo}, {Kawash}, {Linford}, {Molaro}, {Molina}, {Mr{\'o}z}, {Mukai}, {Orio}, {Panurach}, {Senchyna}, {Shappee}, {Shen}, {Sokoloski}, {Sokolovsky}, {Urquhart}, \& {Williams}}]{aydi2023}
{Aydi}, E., {Chomiuk}, L., {Miko{\l}ajewska}, J., {et~al.} 2023, \mnras, 524, 1946, \dodoi{10.1093/mnras/stad1914}

\bibitem[{{Bellm} {et~al.}(2019){Bellm}, {Kulkarni}, {Graham}, {Dekany}, {Smith}, {Riddle}, {Masci}, {Helou}, {Prince}, {Adams}, {Barbarino}, {Barlow}, {Bauer}, {Beck}, {Belicki}, {Biswas}, {Blagorodnova}, {Bodewits}, {Bolin}, {Brinnel}, {Brooke}, {Bue}, {Bulla}, {Burruss}, {Cenko}, {Chang}, {Connolly}, {Coughlin}, {Cromer}, {Cunningham}, {De}, {Delacroix}, {Desai}, {Duev}, {Eadie}, {Farnham}, {Feeney}, {Feindt}, {Flynn}, {Franckowiak}, {Frederick}, {Fremling}, {Gal-Yam}, {Gezari}, {Giomi}, {Goldstein}, {Golkhou}, {Goobar}, {Groom}, {Hacopians}, {Hale}, {Henning}, {Ho}, {Hover}, {Howell}, {Hung}, {Huppenkothen}, {Imel}, {Ip}, {Ivezi{\'c}}, {Jackson}, {Jones}, {Juric}, {Kasliwal}, {Kaspi}, {Kaye}, {Kelley}, {Kowalski}, {Kramer}, {Kupfer}, {Landry}, {Laher}, {Lee}, {Lin}, {Lin}, {Lunnan}, {Giomi}, {Mahabal}, {Mao}, {Miller}, {Monkewitz}, {Murphy}, {Ngeow}, {Nordin}, {Nugent}, {Ofek}, {Patterson}, {Penprase}, {Porter}, {Rauch}, {Rebbapragada}, {Reiley}, {Rigault}, {Rodriguez}, {van Roestel}, {Rusholme}, {van
  Santen}, {Schulze}, {Shupe}, {Singer}, {Soumagnac}, {Stein}, {Surace}, {Sollerman}, {Szkody}, {Taddia}, {Terek}, {Van Sistine}, {van Velzen}, {Vestrand}, {Walters}, {Ward}, {Ye}, {Yu}, {Yan}, \& {Zolkower}}]{bellm2019}
{Bellm}, E.~C., {Kulkarni}, S.~R., {Graham}, M.~J., {et~al.} 2019, \pasp, 131, 018002, \dodoi{10.1088/1538-3873/aaecbe}

\bibitem[{{Bhargava} {et~al.}(2024){Bhargava}, {Dewangan}, {Anupama}, {Kamath}, {Sonith}, {Singh}, {Drake}, {Beardmore}, {Luna}, {Orio}, \& {Page}}]{Bhargava+24}
{Bhargava}, Y., {Dewangan}, G.~C., {Anupama}, G.~C., {et~al.} 2024, \mnras, 528, 28, \dodoi{10.1093/mnras/stad3870}

\bibitem[{{Chomiuk} {et~al.}(2021){Chomiuk}, {Metzger}, \& {Shen}}]{cms2021}
{Chomiuk}, L., {Metzger}, B.~D., \& {Shen}, K.~J. 2021, \araa, 59, 391, \dodoi{10.1146/annurev-astro-112420-114502}

\bibitem[{{Chomiuk} {et~al.}(2014){Chomiuk}, {Linford}, {Yang}, {O'Brien}, {Paragi}, {Mioduszewski}, {Beswick}, {Cheung}, {Mukai}, {Nelson}, {Ribeiro}, {Rupen}, {Sokoloski}, {Weston}, {Zheng}, {Bode}, {Eyres}, {Roy}, \& {Taylor}}]{chomiuk2014}
{Chomiuk}, L., {Linford}, J.~D., {Yang}, J., {et~al.} 2014, \nat, 514, 339, \dodoi{10.1038/nature13773}

\bibitem[{{Corbett} {et~al.}(2023){Corbett}, {Carney}, {Gonzalez}, {Fors}, {Galliher}, {Glazier}, {Howard}, {Law}, {Quimby}, {Ratzloff}, \& {Soto}}]{corbett2023}
{Corbett}, H., {Carney}, J., {Gonzalez}, R., {et~al.} 2023, \apjs, 265, 63, \dodoi{10.3847/1538-4365/acbd41}

\bibitem[{{De Young} \& {Schmidt}(1994)}]{DeYoung&Schmidt94}
{De Young}, J.~A., \& {Schmidt}, R.~E. 1994, \apjl, 431, L47, \dodoi{10.1086/187469}

\bibitem[{{Doherty} {et~al.}(2015){Doherty}, {Gil-Pons}, {Siess}, {Lattanzio}, \& {Lau}}]{doherty2015}
{Doherty}, C.~L., {Gil-Pons}, P., {Siess}, L., {Lattanzio}, J.~C., \& {Lau}, H. H.~B. 2015, \mnras, 446, 2599, \dodoi{10.1093/mnras/stu2180}

\bibitem[{{Drake} {et~al.}(2021){Drake}, {Ness}, {Page}, {Luna}, {Beardmore}, {Orio}, {Osborne}, {Mr{\'o}z}, {Starrfield}, {Banerjee}, {Balman}, {Darnley}, {Bhargava}, {Dewangan}, \& {Singh}}]{Drake+21}
{Drake}, J.~J., {Ness}, J.-U., {Page}, K.~L., {et~al.} 2021, \apjl, 922, L42, \dodoi{10.3847/2041-8213/ac34fd}

\bibitem[{{Eggleton}(1983)}]{eggleton1983}
{Eggleton}, P.~P. 1983, \apj, 268, 368, \dodoi{10.1086/160960}

\bibitem[{{Frank} {et~al.}(2002){Frank}, {King}, \& {Raine}}]{Frank+02}
{Frank}, J., {King}, A., \& {Raine}, D.~J. 2002, {Accretion Power in Astrophysics: Third Edition}

\bibitem[{{Gallagher} \& {Code}(1974)}]{Gallagher&Code74}
{Gallagher}, J.~S., I., \& {Code}, A.~D. 1974, \apj, 189, 303, \dodoi{10.1086/152804}

\bibitem[{{Gallagher} \& {Starrfield}(1978)}]{Gallagher&Starrfield78}
{Gallagher}, J.~S., \& {Starrfield}, S. 1978, \araa, 16, 171, \dodoi{10.1146/annurev.aa.16.090178.001131}

\bibitem[{{Graham} {et~al.}(2019){Graham}, {Kulkarni}, {Bellm}, {Adams}, {Barbarino}, {Blagorodnova}, {Bodewits}, {Bolin}, {Brady}, {Cenko}, {Chang}, {Coughlin}, {De}, {Eadie}, {Farnham}, {Feindt}, {Franckowiak}, {Fremling}, {Gezari}, {Ghosh}, {Goldstein}, {Golkhou}, {Goobar}, {Ho}, {Huppenkothen}, {Ivezi{\'c}}, {Jones}, {Juric}, {Kaplan}, {Kasliwal}, {Kelley}, {Kupfer}, {Lee}, {Lin}, {Lunnan}, {Mahabal}, {Miller}, {Ngeow}, {Nugent}, {Ofek}, {Prince}, {Rauch}, {van Roestel}, {Schulze}, {Singer}, {Sollerman}, {Taddia}, {Yan}, {Ye}, {Yu}, {Barlow}, {Bauer}, {Beck}, {Belicki}, {Biswas}, {Brinnel}, {Brooke}, {Bue}, {Bulla}, {Burruss}, {Connolly}, {Cromer}, {Cunningham}, {Dekany}, {Delacroix}, {Desai}, {Duev}, {Feeney}, {Flynn}, {Frederick}, {Gal-Yam}, {Giomi}, {Groom}, {Hacopians}, {Hale}, {Helou}, {Henning}, {Hover}, {Hillenbrand}, {Howell}, {Hung}, {Imel}, {Ip}, {Jackson}, {Kaspi}, {Kaye}, {Kowalski}, {Kramer}, {Kuhn}, {Landry}, {Laher}, {Mao}, {Masci}, {Monkewitz}, {Murphy}, {Nordin}, {Patterson}, {Penprase},
  {Porter}, {Rebbapragada}, {Reiley}, {Riddle}, {Rigault}, {Rodriguez}, {Rusholme}, {van Santen}, {Shupe}, {Smith}, {Soumagnac}, {Stein}, {Surace}, {Szkody}, {Terek}, {Van Sistine}, {van Velzen}, {Vestrand}, {Walters}, {Ward}, {Zhang}, \& {Zolkower}}]{graham2019}
{Graham}, M.~J., {Kulkarni}, S.~R., {Bellm}, E.~C., {et~al.} 2019, \pasp, 131, 078001, \dodoi{10.1088/1538-3873/ab006c}

\bibitem[{{Habtie} {et~al.}(2024){Habtie}, {Das}, {Pandey}, {Ashok}, \& {Dubovsky}}]{habtie2024}
{Habtie}, G.~R., {Das}, R., {Pandey}, R., {Ashok}, N.~M., \& {Dubovsky}, P.~A. 2024, \mnras, 527, 1405, \dodoi{10.1093/mnras/stad3295}

\bibitem[{{Hachisu} \& {Kato}(2022)}]{hachisu_kato2022}
{Hachisu}, I., \& {Kato}, M. 2022, \apj, 939, 1, \dodoi{10.3847/1538-4357/ac9475}

\bibitem[{{Hillman} {et~al.}(2014){Hillman}, {Prialnik}, {Kovetz}, {Shara}, \& {Neill}}]{hil14}
{Hillman}, Y., {Prialnik}, D., {Kovetz}, A., {Shara}, M.~M., \& {Neill}, J.~D. 2014, \mnras, 437, 1962, \dodoi{10.1093/mnras/stt2027}

\bibitem[{{Hounsell} {et~al.}(2010){Hounsell}, {Bode}, {Hick}, {Buffington}, {Jackson}, {Clover}, {Shafter}, {Darnley}, {Mawson}, {Steele}, {Evans}, {Eyres}, \& {O'Brien}}]{hou10}
{Hounsell}, R., {Bode}, M.~F., {Hick}, P.~P., {et~al.} 2010, \apj, 724, 480, \dodoi{10.1088/0004-637X/724/1/480}

\bibitem[{{Kato} \& {Hachisu}(1994)}]{kato_hachisu1994}
{Kato}, M., \& {Hachisu}, I. 1994, \apj, 437, 802, \dodoi{10.1086/175041}

\bibitem[{{Kato} {et~al.}(2024){Kato}, {Saio}, \& {Hachisu}}]{ksh2024}
{Kato}, M., {Saio}, H., \& {Hachisu}, I. 2024, \pasj, 76, 666, \dodoi{10.1093/pasj/psae038}

\bibitem[{{{Kloppenborg}, B.~K.}(2023)}]{aavso}
{{Kloppenborg}, B.~K.} 2023, Observations from the AAVSO International Database, \url{https://www.aavso.org}

\bibitem[{{Knigge}(2006)}]{Knigge06}
{Knigge}, C. 2006, \mnras, 373, 484, \dodoi{10.1111/j.1365-2966.2006.11096.x}

\bibitem[{{Kochanek} {et~al.}(2017){Kochanek}, {Shappee}, {Stanek}, {Holoien}, {Thompson}, {Prieto}, {Dong}, {Shields}, {Will}, {Britt}, {Perzanowski}, \& {Pojma{\'n}ski}}]{Kochanek+17}
{Kochanek}, C.~S., {Shappee}, B.~J., {Stanek}, K.~Z., {et~al.} 2017, \pasp, 129, 104502, \dodoi{10.1088/1538-3873/aa80d9}

\bibitem[{{K{\"o}nig} {et~al.}(2022){K{\"o}nig}, {Wilms}, {Arcodia}, {Dauser}, {Dennerl}, {Doroshenko}, {Haberl}, {H{\"a}mmerich}, {Kirsch}, {Kreykenbohm}, {Lorenz}, {Malyali}, {Merloni}, {Rau}, {Rauch}, {Sala}, {Schwope}, {Suleimanov}, {Weber}, \& {Werner}}]{konig2022}
{K{\"o}nig}, O., {Wilms}, J., {Arcodia}, R., {et~al.} 2022, \nat, 605, 248, \dodoi{10.1038/s41586-022-04635-y}

\bibitem[{{Law} {et~al.}(2014){Law}, {Fors}, {Wulfken}, {Ratzloff}, \& {Kavanaugh}}]{law14}
{Law}, N.~M., {Fors}, O., {Wulfken}, P., {Ratzloff}, J., \& {Kavanaugh}, D. 2014, in Society of Photo-Optical Instrumentation Engineers (SPIE) Conference Series, Vol. 9145, Ground-based and Airborne Telescopes V, ed. L.~M. {Stepp}, R.~{Gilmozzi}, \& H.~J. {Hall}, 91450Z, \dodoi{10.1117/12.2057031}

\bibitem[{{Law} {et~al.}(2022){Law}, {Corbett}, {Galliher}, {Gonzalez}, {Vasquez}, {Walters}, {Machia}, {Ratzloff}, {Ackley}, {Bizon}, {Clemens}, {Cox}, {Eikenberry}, {Howard}, {Glazier}, {Mann}, {Quimby}, {Reichart}, \& {Trilling}}]{law2022}
{Law}, N.~M., {Corbett}, H., {Galliher}, N.~W., {et~al.} 2022, \pasp, 134, 035003, \dodoi{10.1088/1538-3873/ac4811}

\bibitem[{{Li} {et~al.}(2017){Li}, {Metzger}, {Chomiuk}, {Vurm}, {Strader}, {Finzell}, {Beloborodov}, {Nelson}, {Shappee}, {Kochanek}, {Prieto}, {Kafka}, {Holoien}, {Thompson}, {Luckas}, \& {Itoh}}]{Li2017}
{Li}, K.-L., {Metzger}, B.~D., {Chomiuk}, L., {et~al.} 2017, Nature Astronomy, 1, 697, \dodoi{10.1038/s41550-017-0222-1}

\bibitem[{{Lu} {et~al.}(2023){Lu}, {Fuller}, {Quataert}, \& {Bonnerot}}]{lu2023}
{Lu}, W., {Fuller}, J., {Quataert}, E., \& {Bonnerot}, C. 2023, \mnras, 519, 1409, \dodoi{10.1093/mnras/stac3621}

\bibitem[{{Masci} {et~al.}(2023){Masci}, {Laher}, {Rusholme}, {Shupe}, {Paladini}, {Groom}, {Wold}, {Miller}, \& {Drake}}]{masci2023}
{Masci}, F.~J., {Laher}, R.~R., {Rusholme}, B., {et~al.} 2023, arXiv e-prints, arXiv:2305.16279, \dodoi{10.48550/arXiv.2305.16279}

\bibitem[{{Metzger} {et~al.}(2014){Metzger}, {Hasco{\"e}t}, {Vurm}, {Beloborodov}, {Chomiuk}, {Sokoloski}, \& {Nelson}}]{Metzger+14}
{Metzger}, B.~D., {Hasco{\"e}t}, R., {Vurm}, I., {et~al.} 2014, \mnras, 442, 713, \dodoi{10.1093/mnras/stu844}

\bibitem[{{Mroz} {et~al.}(2021){Mroz}, {Burdge}, {Roestel}, {Prince}, {Kong}, \& {Li}}]{mro21}
{Mroz}, P., {Burdge}, K., {Roestel}, J.~v., {et~al.} 2021, The Astronomer's Telegram, 14720, 1

\bibitem[{{Mukai} \& {Pretorius}(2023)}]{mukai2023}
{Mukai}, K., \& {Pretorius}, M.~L. 2023, \mnras, 523, 3192, \dodoi{10.1093/mnras/stad1603}

\bibitem[{{Munari} {et~al.}(2017){Munari}, {Hambsch}, \& {Frigo}}]{munrai2017}
{Munari}, U., {Hambsch}, F.~J., \& {Frigo}, A. 2017, \mnras, 469, 4341, \dodoi{10.1093/mnras/stx1116}

\bibitem[{{Munari} {et~al.}(2020){Munari}, {Moretti}, \& {Maitan}}]{Munari+20}
{Munari}, U., {Moretti}, S., \& {Maitan}, A. 2020, \aap, 639, L10, \dodoi{10.1051/0004-6361/202038403}

\bibitem[{{Munari} {et~al.}(2021){Munari}, {Valisa}, \& {Dallaporta}}]{mun21}
{Munari}, U., {Valisa}, P., \& {Dallaporta}, S. 2021, The Astronomer's Telegram, 14704, 1

\bibitem[{{Murphy-Glaysher} {et~al.}(2022){Murphy-Glaysher}, {Darnley}, {Harvey}, {Newsam}, {Page}, {Starrfield}, {Wagner}, {Woodward}, {Terndrup}, {Kafka}, {Arranz Heras}, {Berardi}, {Bertrand}, {Biernikowicz}, {Boussin}, {Boyd}, {Buchet}, {Bundas}, {Coulter}, {Dejean}, {Diepvens}, {Dvorak}, {Edlin}, {Eenmae}, {Eggenstein}, {Fournier}, {Garde}, {Gout}, {Janzen}, {Jordanov}, {Kiiskinen}, {Lane}, {Larochelle}, {Leadbeater}, {Mankel}, {Martineau}, {Miller}, {Modic}, {Montier}, {Morales Aimar}, {Muyllaert}, {Naves Nogues}, {O'Keeffe}, {Oksanen}, {Pyatnytskyy}, {Rast}, {Rodgers}, {Rodriguez Perez}, {Schorr}, {Schwendeman}, {Shadick}, {Sharpe}, {Sold{\'a}n Alfaro}, {Sove}, {Stone}, {Tordai}, {Venne}, {Vollmann}, {Vrastak}, \& {Wenzel}}]{Murphy-Glaysher+22}
{Murphy-Glaysher}, F.~J., {Darnley}, M.~J., {Harvey}, {\'E}.~J., {et~al.} 2022, \mnras, 514, 6183, \dodoi{10.1093/mnras/stac1577}

\bibitem[{{Patterson}(1984)}]{Patterson84}
{Patterson}, J. 1984, \apjs, 54, 443, \dodoi{10.1086/190940}

\bibitem[{{Patterson}(1994)}]{patterson94}
---. 1994, \pasp, 106, 209, \dodoi{10.1086/133375}

\bibitem[{{Patterson} {et~al.}(2022){Patterson}, {Enenstein}, {de Miguel}, {Epstein-Martin}, {Kemp}, {Sabo}, {Cooney}, {Vanmunster}, {Dubovsky}, {Hambsch}, {Myers}, {Lemay}, {Sokolovsky}, {Collins}, {Campbell}, {Roberts}, {Richmond}, {Brincat}, {Ulowetz}, {Dvorak}, {Tordai}, {Dufoer}, {Cahaly}, {Galdies}, {Goff}, {Wilkin}, \& {Wood}}]{Patterson+22}
{Patterson}, J., {Enenstein}, J., {de Miguel}, E., {et~al.} 2022, \apjl, 940, L56, \dodoi{10.3847/2041-8213/ac9ebe}

\bibitem[{{Pejcha} {et~al.}(2016){Pejcha}, {Metzger}, \& {Tomida}}]{pejcha2016}
{Pejcha}, O., {Metzger}, B.~D., \& {Tomida}, K. 2016, \mnras, 455, 4351, \dodoi{10.1093/mnras/stv2592}

\bibitem[{{Prialnik} {et~al.}(1979){Prialnik}, {Shara}, \& {Shaviv}}]{Prialnik+79}
{Prialnik}, D., {Shara}, M.~M., \& {Shaviv}, G. 1979, \aap, 72, 192

\bibitem[{{Quataert} {et~al.}(2016){Quataert}, {Fern{\'a}ndez}, {Kasen}, {Klion}, \& {Paxton}}]{quataert2016}
{Quataert}, E., {Fern{\'a}ndez}, R., {Kasen}, D., {Klion}, H., \& {Paxton}, B. 2016, \mnras, 458, 1214, \dodoi{10.1093/mnras/stw365}

\bibitem[{{Quimby} {et~al.}(2021){Quimby}, {Shafter}, \& {Corbett}}]{quimby2021}
{Quimby}, R.~M., {Shafter}, A.~W., \& {Corbett}, H. 2021, Research Notes of the American Astronomical Society, 5, 160, \dodoi{10.3847/2515-5172/ac14c0}

\bibitem[{{Schaefer}(2021)}]{schaefer_AAS2021}
{Schaefer}, B.~E. 2021, in American Astronomical Society Meeting Abstracts, Vol. 238, American Astronomical Society Meeting Abstracts, 225.03

\bibitem[{{Schaefer}(2022{\natexlab{a}})}]{Schaefer22b}
{Schaefer}, B.~E. 2022{\natexlab{a}}, \mnras, 517, 3640, \dodoi{10.1093/mnras/stac2089}

\bibitem[{{Schaefer}(2022{\natexlab{b}})}]{schaefer2022}
---. 2022{\natexlab{b}}, \mnras, 517, 6150, \dodoi{10.1093/mnras/stac2900}

\bibitem[{{Schaefer} {et~al.}(2013){Schaefer}, {Landolt}, {Linnolt}, {Stubbings}, {Pojmanski}, {Plummer}, {Kerr}, {Nelson}, {Carstens}, {Streamer}, {Richards}, {Myers}, \& {Dillon}}]{schaefer2013}
{Schaefer}, B.~E., {Landolt}, A.~U., {Linnolt}, M., {et~al.} 2013, \apj, 773, 55, \dodoi{10.1088/0004-637X/773/1/55}

\bibitem[{{Schmidt} {et~al.}(2021){Schmidt}, {Shugarov}, \& {Afonina}}]{schmidt2021}
{Schmidt}, R.~E., {Shugarov}, S.~Y., \& {Afonina}, M.~D. 2021, \jaavso, 49, 257

\bibitem[{{Shappee} {et~al.}(2014){Shappee}, {Prieto}, {Grupe}, {Kochanek}, {Stanek}, {De Rosa}, {Mathur}, {Zu}, {Peterson}, {Pogge}, {Komossa}, {Im}, {Jencson}, {Holoien}, {Basu}, {Beacom}, {Szczygie{\l}}, {Brimacombe}, {Adams}, {Campillay}, {Choi}, {Contreras}, {Dietrich}, {Dubberley}, {Elphick}, {Foale}, {Giustini}, {Gonzalez}, {Hawkins}, {Howell}, {Hsiao}, {Koss}, {Leighly}, {Morrell}, {Mudd}, {Mullins}, {Nugent}, {Parrent}, {Phillips}, {Pojmanski}, {Rosing}, {Ross}, {Sand}, {Terndrup}, {Valenti}, {Walker}, \& {Yoon}}]{Shappee+14}
{Shappee}, B.~J., {Prieto}, J.~L., {Grupe}, D., {et~al.} 2014, \apj, 788, 48, \dodoi{10.1088/0004-637X/788/1/48}

\bibitem[{{Shen} \& {Quataert}(2022)}]{shen_quataert2022}
{Shen}, K.~J., \& {Quataert}, E. 2022, \apj, 938, 31, \dodoi{10.3847/1538-4357/ac9136}

\bibitem[{{Shugarov} \& {Afonina}(2021)}]{shugarov21}
{Shugarov}, S., \& {Afonina}, M. 2021, The Astronomer's Telegram, 14835, 1

\bibitem[{{Sokolovsky} {et~al.}(2023){Sokolovsky}, {Johnson}, {Buson}, {Jean}, {Cheung}, {Mukai}, {Chomiuk}, {Aydi}, {Molina}, {Kawash}, {Linford}, {Mioduszewski}, {Rupen}, {Sokoloski}, {Williams}, {Steinberg}, {Vurm}, {Metzger}, {Page}, {Orio}, {Quimby}, {Shafter}, {Corbett}, {Bolzoni}, {DeYoung}, {Menzies}, {Romanov}, {Richmond}, {Ulowetz}, {Vanmunster}, {Williamson}, {Lane}, {Bartnik}, {Bellaver}, {Bruinsma}, {Dugan}, {Fedewa}, {Gerhard}, {Painter}, {Peterson}, {Rodriguez}, {Smith}, {Sullivan}, \& {Watson}}]{sokolovsky2023}
{Sokolovsky}, K.~V., {Johnson}, T.~J., {Buson}, S., {et~al.} 2023, \mnras, 521, 5453, \dodoi{10.1093/mnras/stad887}

\bibitem[{{Starrfield} {et~al.}(1990){Starrfield}, {Truran}, {Sparks}, {Krautter}, \& {MacDonald}}]{starrfield1990}
{Starrfield}, S., {Truran}, J.~W., {Sparks}, W.~M., {Krautter}, J., \& {MacDonald}, J. 1990, in IAU Colloq. 122: Physics of Classical Novae, ed. A.~{Cassatella} \& R.~{Viotti}, Vol. 369, 306, \dodoi{10.1007/3-540-53500-4_143}

\bibitem[{{Strope} {et~al.}(2010){Strope}, {Schaefer}, \& {Henden}}]{2010AJ....140...34S}
{Strope}, R.~J., {Schaefer}, B.~E., \& {Henden}, A.~A. 2010, \aj, 140, 34, \dodoi{10.1088/0004-6256/140/1/34}

\bibitem[{{Thompson}(2017)}]{thompson2017}
{Thompson}, W.~T. 2017, \mnras, 470, 4061, \dodoi{10.1093/mnras/stx1552}

\bibitem[{{Tonry} {et~al.}(2018){Tonry}, {Denneau}, {Flewelling}, {Heinze}, {Onken}, {Smartt}, {Stalder}, {Weiland}, \& {Wolf}}]{tonry2018}
{Tonry}, J.~L., {Denneau}, L., {Flewelling}, H., {et~al.} 2018, \apj, 867, 105, \dodoi{10.3847/1538-4357/aae386}

\bibitem[{{van Paradijs}(1983)}]{vanparadijs83}
{van Paradijs}, J. 1983, in Accretion-Driven Stellar X-ray Sources, ed. W.~H.~G. {Lewin} \& E.~P.~J. {van den Heuvel}, 189--260

\bibitem[{{van Paradijs} \& {McClintock}(1994)}]{vanparadijs&mcclintock94}
{van Paradijs}, J., \& {McClintock}, J.~E. 1994, \aap, 290, 133

\bibitem[{{Wagner} {et~al.}(2021){Wagner}, {Woodward}, {Starrfield}, {Banerjee}, \& {Evans}}]{wag21}
{Wagner}, R.~M., {Woodward}, C.~E., {Starrfield}, S., {Banerjee}, D.~P.~K., \& {Evans}, A. 2021, The Astronomer's Telegram, 14746, 1

\bibitem[{{Woodward} {et~al.}(2021{\natexlab{a}}){Woodward}, {Banerjee}, {Geballe}, {Page}, {Starrfield}, \& {Wagner}}]{woodward21}
{Woodward}, C.~E., {Banerjee}, D.~P.~K., {Geballe}, T.~R., {et~al.} 2021{\natexlab{a}}, \apjl, 922, L10, \dodoi{10.3847/2041-8213/ac3518}

\bibitem[{{Woodward} {et~al.}(2021{\natexlab{b}}){Woodward}, {Wagner}, {Starrfield}, {Kumar}, {Srivastava}, {Banerjee}, {Joshi}, {IIyin}, {Strassmeier}, \& {Evans}}]{woodward2021ATel}
{Woodward}, C.~E., {Wagner}, R.~M., {Starrfield}, S., {et~al.} 2021{\natexlab{b}}, The Astronomer's Telegram, 14723, 1

\end{thebibliography}
\bibliographystyle{aasjournal}

\appendix

\section{Peak Magnitude and Decline Rate of V1674 Her}\label{sec:t2}

\begin{figure}
    \centering
    \includegraphics[width=\columnwidth]{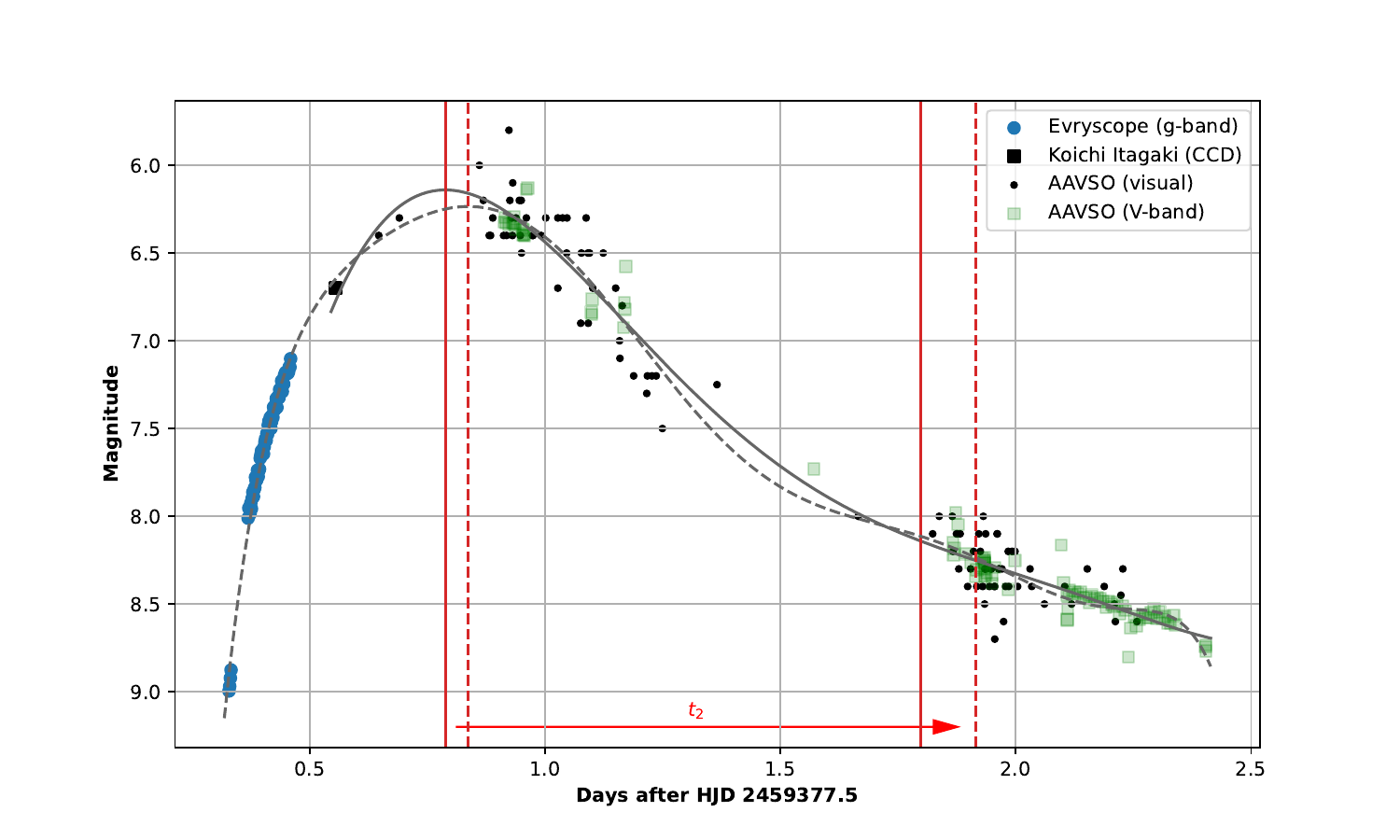}
    \caption{Polynomial fits to the AAVSO data (grey solid curve) and these data supplemented with the faster-rise portion of the Evryscope $g$-band light curve (dashed grey curve) suggest similar peak magnitudes ($V_{\rm max} \sim 6.2$\,mag) and $t_2$ times of 1.01\,d to 1.08\,d.
    }
    \label{fig:t2}
\end{figure}

Here we use data from the AAVSO \citep{aavso} to determine the peak V-band magnitude of \nova\ and $t_2$, the time it took to fade to 2\,mag below this peak. Figure~\ref{fig:t2} shows visual and $V$-band measurements of of \nova\ around its peak along with an unfiltered CCD measurement from Koichi Itagaki and the final, faster-rise section of the Evryscope $g$-band light curve for reference. The solid grey curve is a 6th order polynomial fit excluding the Evryscope data (we assume $V$-band and visual magnitudes to be equivalent at these phases), and the dashed grey curve is a 10th order polynomial fit \textit{including} the Evryscope data (this further assumes $g-V=0$). These fits predict peak magnitudes of $V=6.14$\,mag at $t_{\rm ref} + 0.7878$\,d and $V=6.23$\,mag at $t_{\rm ref} + 0.8371$\,d, respectively. The $t_2$ values for these fits are 1.01\,d and 1.08\,d, respectively. We thus conclude that \nova\ reached a peak $V$-band magnitude of $V=6.19 \pm 0.05$ and declined by 2 magnitudes in $1.04 \pm 0.03$\,d.

\startlongtable
\begin{deluxetable}{lrrrrr}
\tabletypesize{\scriptsize}
\tablewidth{0pt}
\tablecaption{ASAS-SN Photometry \label{tab:asas-sn}}
\tablehead{
\colhead{HJD} & \colhead{Flux}& \colhead{$\sigma_{\rm Flux}$} & \colhead{$g$} & \colhead{$\sigma_g$} & \colhead{Limit} \\
\colhead{} & \colhead{mJy} & \colhead{mJy} & \colhead{mag} & \colhead{mag} & \colhead{mag} \\} 
\colnumbers
\startdata
2459377.6944 & 0.557 & 0.137 & 17.035 & 0.267 & 16.810 \\
2459377.6956 & 1.161 & 0.153 & 16.238 & 0.143 & 16.693 \\
2459377.6968 & 0.978 & 0.143 & 16.424 & 0.159 & 16.762 \\
\enddata
\tablecomments{HJD values mark the beginning of the 90\,s exposures. Limits are $5\sigma$.}
\end{deluxetable}

\startlongtable
\begin{deluxetable}{lrrrrr}
\tabletypesize{\scriptsize}
\tablewidth{0pt}
\tablecaption{Evryscope Photometry \label{tab:photometry}}
\tablehead{
\colhead{HJD} & \colhead{Flux}& \colhead{$\sigma_{\rm Flux}$} & \colhead{$g$} & \colhead{$\sigma_g$} & \colhead{Limit} \\
\colhead{} & \colhead{mJy} & \colhead{mJy} & \colhead{mag} & \colhead{mag} & \colhead{mag} \\} 
\colnumbers
\startdata
2459377.7527 &      2.52 &      1.08 & \nodata & \nodata & 15.064 \\
2459377.7541 &      2.22 &      1.08 & \nodata & \nodata & 15.094 \\
2459377.7556 &      2.57 &      1.07 & \nodata & \nodata & 15.133 \\
2459377.7570 &      2.61 &      1.11 & \nodata & \nodata & 15.149 \\
2459377.7584 &      4.19 &      1.11 &  14.844 &   0.288 & 15.114 \\
2459377.7598 &      3.64 &      1.10 &  14.997 &   0.328 & 15.140 \\
2459377.7612 &      4.72 &      1.09 &  14.715 &   0.250 & 15.128 \\
2459377.7627 &      3.30 &      1.08 &  15.104 &   0.356 & 15.123 \\
2459377.7641 &      4.96 &      1.09 &  14.662 &   0.238 & 15.203 \\
2459377.7655 &      3.67 &      1.05 &  14.988 &   0.310 & 15.092 \\
2459377.7670 &      5.28 &      1.07 &  14.594 &   0.220 & 15.088 \\
2459377.7684 &      5.38 &      1.08 &  14.573 &   0.217 & 15.121 \\
2459377.7698 &      4.54 &      1.08 &  14.757 &   0.258 & 15.086 \\
2459377.7712 &      4.93 &      1.10 &  14.667 &   0.242 & 15.097 \\
2459377.7726 &      7.62 &      1.11 &  14.195 &   0.158 & 15.147 \\
2459377.7741 &      5.50 &      1.11 &  14.548 &   0.218 & 15.137 \\
2459377.7755 &      3.74 &      1.09 &  14.967 &   0.316 & 15.087 \\
2459377.7769 &      1.17 &      1.06 & \nodata & \nodata & 15.144 \\
2459377.7784 &      4.50 &      1.06 &  14.767 &   0.256 & 15.243 \\
2459377.7798 &      4.67 &      1.06 &  14.727 &   0.247 & 15.129 \\
2459377.7812 &     10.86 &      1.07 &  13.811 &   0.107 & 15.162 \\
2459377.7826 &     19.12 &      1.11 &  13.196 &   0.063 & 15.121 \\
2459377.7840 &     34.61 &      1.17 &  12.552 &   0.037 & 15.214 \\
2459377.7855 &     53.06 &      1.21 &  12.088 &   0.025 & 15.189 \\
2459377.7869 &     74.93 &      1.27 &  11.713 &   0.018 & 15.212 \\
2459377.7883 &    101.63 &      1.35 &  11.382 &   0.014 & 15.227 \\
2459377.7897 &    124.83 &      1.39 &  11.159 &   0.012 & 15.190 \\
2459377.7912 &    157.23 &      1.45 &  10.909 &   0.010 & 15.205 \\
2459377.7926 &    178.65 &      1.48 &  10.770 &   0.009 & 15.254 \\
2459377.7940 &    206.70 &      1.53 &  10.612 &   0.008 & 15.211 \\
2459377.7954 &    233.40 &      1.56 &  10.480 &   0.007 & 15.155 \\
2459377.7969 &    268.33 &      1.61 &  10.328 &   0.007 & 15.172 \\
2459377.7983 &    290.98 &      1.66 &  10.240 &   0.006 & 15.161 \\
2459377.7997 &    322.49 &      1.73 &  10.129 &   0.006 & 15.201 \\
2459377.8011 &    351.36 &      1.78 &  10.036 &   0.006 & 15.163 \\
2459377.8026 &    372.68 &      1.83 &   9.972 &   0.005 & 15.218 \\
2459377.8040 &    399.69 &      1.88 &   9.896 &   0.005 & 15.244 \\
2459377.8054 &    430.99 &      1.92 &   9.814 &   0.005 & 15.168 \\
2459377.8068 &    456.27 &      2.01 &   9.752 &   0.005 & 15.167 \\
2459377.8083 &    484.00 &      2.07 &   9.688 &   0.005 & 15.156 \\
2459377.8097 &    511.76 &      2.11 &   9.627 &   0.004 & 15.249 \\
2459377.8111 &    545.85 &      2.16 &   9.557 &   0.004 & 15.182 \\
2459377.8125 &    574.85 &      2.32 &   9.501 &   0.004 & 15.177 \\
2459377.8140 &    605.42 &      2.36 &   9.445 &   0.004 & 15.261 \\
2459377.8154 &    640.89 &      2.40 &   9.383 &   0.004 & 15.215 \\
2459377.8168 &    665.59 &      2.43 &   9.342 &   0.004 & 15.214 \\
2459377.8183 &    702.12 &      2.55 &   9.284 &   0.004 & 15.275 \\
2459377.8197 &    730.93 &      2.65 &   9.240 &   0.004 & 15.151 \\
2459377.8211 &    753.47 &      2.75 &   9.207 &   0.004 & 15.174 \\
2459377.8225 &    785.42 &      2.78 &   9.162 &   0.004 & 15.155 \\
2459377.8239 &    803.56 &      2.79 &   9.137 &   0.004 & 15.128 \\
2459377.8254 &    841.10 &      2.83 &   9.088 &   0.004 & 15.236 \\
2459377.8268 &    865.52 &      3.04 &   9.057 &   0.004 & 15.217 \\
2459377.8282 &    916.09 &      2.99 &   8.995 &   0.004 & 15.208 \\
2459377.8296 &    938.54 &      3.10 &   8.969 &   0.004 & 15.191 \\
2459377.8311 &    980.05 &      3.38 &   8.922 &   0.004 & 15.199 \\
2459377.8325 &   1022.46 &      3.44 &   8.876 &   0.004 & 15.168 \\
2459377.8688 &   2268.52 &      8.77 &   8.011 &   0.004 & 15.795 \\
2459377.8702 &   2390.97 &     10.16 &   7.954 &   0.005 & 15.792 \\
2459377.8716 &   2389.91 &      9.52 &   7.954 &   0.004 & 15.871 \\
2459377.8731 &   2319.60 &      9.28 &   7.986 &   0.004 & 15.863 \\
2459377.8745 &   2454.78 &      9.72 &   7.925 &   0.004 & 15.820 \\
2459377.8759 &   2383.21 &      9.20 &   7.957 &   0.004 & 15.751 \\
2459377.8773 &   2516.92 &     10.03 &   7.898 &   0.004 & 15.800 \\
2459377.8788 &   2599.60 &     10.29 &   7.863 &   0.004 & 15.764 \\
2459377.8802 &   2538.88 &      9.46 &   7.888 &   0.004 & 15.792 \\
2459377.8816 &   2635.60 &     10.32 &   7.848 &   0.004 & 15.843 \\
2459377.8830 &   2663.69 &     10.04 &   7.836 &   0.004 & 15.791 \\
2459377.8845 &   2812.66 &     10.75 &   7.777 &   0.004 & 15.803 \\
2459377.8859 &   2758.41 &     10.33 &   7.798 &   0.004 & 15.825 \\
2459377.8873 &   2778.49 &     10.72 &   7.790 &   0.004 & 15.846 \\
2459377.8887 &   2917.14 &     11.76 &   7.738 &   0.004 & 15.822 \\
2459377.8902 &   2821.56 &      9.73 &   7.774 &   0.004 & 15.718 \\
2459377.8916 &   2911.41 &     10.91 &   7.740 &   0.004 & 15.699 \\
2459377.8930 &   2934.27 &     10.35 &   7.731 &   0.004 & 15.788 \\
2459377.8944 &   3106.28 &     11.32 &   7.669 &   0.004 & 15.786 \\
2459377.8959 &   3163.45 &     11.56 &   7.650 &   0.004 & 15.738 \\
2459377.8973 &   3229.46 &     11.89 &   7.627 &   0.004 & 15.775 \\
2459377.8987 &   3166.37 &     11.06 &   7.649 &   0.004 & 15.784 \\
2459377.9001 &   3179.87 &     11.14 &   7.644 &   0.004 & 15.764 \\
2459377.9016 &   3178.30 &     11.18 &   7.645 &   0.004 & 15.749 \\
2459377.9030 &   3297.47 &     11.84 &   7.605 &   0.004 & 15.739 \\
2459377.9044 &   3382.01 &     12.01 &   7.577 &   0.004 & 15.788 \\
2459377.9058 &   3441.23 &     12.15 &   7.558 &   0.004 & 15.740 \\
2459377.9073 &   3410.75 &     11.96 &   7.568 &   0.004 & 15.801 \\
2459377.9087 &   3538.72 &     12.58 &   7.528 &   0.004 & 15.795 \\
2459377.9101 &   3521.68 &     12.21 &   7.533 &   0.004 & 15.834 \\
2459377.9115 &   3696.03 &     12.62 &   7.481 &   0.004 & 15.825 \\
2459377.9130 &   3775.92 &     12.65 &   7.457 &   0.004 & 15.697 \\
2459377.9144 &   3767.51 &     12.83 &   7.460 &   0.004 & 15.748 \\
2459377.9158 &   3847.25 &     12.73 &   7.437 &   0.004 & 15.690 \\
2459377.9172 &   3631.40 &     11.72 &   7.500 &   0.004 & 15.740 \\
2459377.9187 &   3764.39 &     11.78 &   7.461 &   0.003 & 15.745 \\
2459377.9201 &   3897.85 &     13.42 &   7.423 &   0.004 & 15.778 \\
2459377.9215 &   3849.65 &     12.33 &   7.436 &   0.003 & 15.673 \\
2459377.9229 &   4062.71 &     13.83 &   7.378 &   0.004 & 15.753 \\
2459377.9244 &   4045.05 &     12.79 &   7.383 &   0.003 & 15.780 \\
2459377.9258 &   4013.44 &     13.00 &   7.391 &   0.004 & 15.726 \\
2459377.9272 &   4117.75 &     13.59 &   7.363 &   0.004 & 15.753 \\
2459377.9286 &   4248.82 &     13.87 &   7.329 &   0.004 & 15.778 \\
2459377.9301 &   4055.42 &     12.23 &   7.380 &   0.003 & 15.805 \\
2459377.9315 &   4265.78 &     13.43 &   7.325 &   0.003 & 15.802 \\
2459377.9329 &   4301.47 &     13.43 &   7.316 &   0.003 & 15.699 \\
2459377.9343 &   4265.11 &     12.93 &   7.325 &   0.003 & 15.763 \\
2459377.9358 &   4454.89 &     13.70 &   7.278 &   0.003 & 15.796 \\
2459377.9372 &   4459.74 &     13.14 &   7.277 &   0.003 & 15.698 \\
2459377.9386 &   4414.00 &     12.90 &   7.288 &   0.003 & 15.724 \\
2459377.9400 &   4661.62 &     14.22 &   7.229 &   0.003 & 15.703 \\
2459377.9415 &   4407.32 &     13.04 &   7.290 &   0.003 & 15.772 \\
2459377.9429 &   4588.03 &     13.48 &   7.246 &   0.003 & 15.765 \\
2459377.9443 &   4584.41 &     13.39 &   7.247 &   0.003 & 15.800 \\
2459377.9457 &   4785.28 &     14.14 &   7.200 &   0.003 & 15.781 \\
2459377.9472 &   4832.53 &     14.57 &   7.190 &   0.003 & 15.739 \\
2459377.9486 &   4876.00 &     13.70 &   7.180 &   0.003 & 15.703 \\
2459377.9535 &   4852.54 &     12.50 &   7.185 &   0.003 & 15.462 \\
2459377.9549 &   4894.31 &     12.61 &   7.176 &   0.003 & 15.460 \\
2459377.9563 &   5018.83 &     12.87 &   7.148 &   0.003 & 15.439 \\
2459377.9577 &   5010.01 &     12.74 &   7.150 &   0.003 & 15.428 \\
2459377.9591 &   5237.22 &     13.53 &   7.102 &   0.003 & 15.408 \\
\enddata
\tablecomments{HJD values mark the midpoint of the 120\,s exposures. Uncertainties are statistical only. Limits are for 50\% completeness. }
\end{deluxetable}

\end{document}